\documentclass[preprintnumbers,superscriptaddress,noshowpacs,eqsecnum,twocolumn,prd]{revtex4}

\usepackage{graphicx}
\usepackage{amsmath,amsthm,amssymb}
\usepackage{mathrsfs,bbm}

\providecommand{\realnums}{\mathbbm{R}}
\providecommand{\abs}[1]{\left\lvert {#1} \right\rvert} 
\providecommand{\nvec}[1]{\boldsymbol{#1}}              
\providecommand{\ud}[0]{\mathrm{d}}                     

\DeclareMathOperator{\cohull}{co}
\DeclareMathOperator{\cohullcl}{\overline{co}} 
\DeclareMathOperator{\cocone}{K}
\DeclareMathOperator{\coconecl}{\overline{K}}  

\providecommand{\sigmadip}{\hat{\sigma}}
\providecommand{\sigmatot}{\sigma}
\providecommand{\sigmared}{\sigma_{\mathrm{r}}}

\begin{document}

\preprint{ECT$^*$-07-15, HD-THEP-07-17}
\title{On the Range of Validity of the Dipole Picture}

\author{Carlo Ewerz}
\email{C.Ewerz@thphys.uni-heidelberg.de}
\affiliation{ECT\,$^*$,
     Strada delle Tabarelle 286, I-38050 Villazzano (Trento), Italy}

\author{Andreas von Manteuffel}
\email{A.v.Manteuffel@thphys.uni-heidelberg.de}

\author{Otto Nachtmann}
\email{O.Nachtmann@thphys.uni-heidelberg.de}
\affiliation{Institut f\"ur Theoretische Physik, Universit\"at Heidelberg,
    Philosophenweg 16, D-69120 Heidelberg, Germany}

\begin{abstract}
\vspace*{1cm}
\noindent
Ein Photon kommt recht schnell daher \\
und glaubt, da{\ss} es ein Dipol w\"ar. \\
Ein Proton aus recht gro{\ss}er Ferne \\
sieht dies nur allzu gerne. \\
Und was die zwei dann treiben, \\
kann keine Theorie beschreiben,\\
Gluonen fliegen hin und her, \\
das versteht kein Mensch nicht mehr.\\
Viel komplizierter als gedacht\\
ist doch die QCD gemacht. \\
Schon die Sache mit dem Dipol war -- \\
so ist es wohl -- nicht ganz so klar. \\
Der Ph\"anomenologe, keine Frage,\\
inspiziert trotzdem die Datenlage.\\
Modelle sind dann schnell zur Hand, \\
vielleicht sogar mit Fehlerband. \\
Und kommt die Kurve an die Punkte ran, \\
weht niemanden ein Zweifel an.\\
Auf da{\ss} das nicht so bleiben soll,\\
schreibt unsereins dies Paper voll.\\[.5cm]

We derive correlated bounds on ratios of deep inelastic 
structure functions from the dipole picture of 
photon-hadron scattering at high energies. In particular we 
consider ratios of the longitudinal structure function, the total structure 
function and the charm part of the latter. We also 
consider ratios of total structure functions taken at 
the same energy but at three different photon 
virtualities. It is shown that by confronting these bounds with 
experimental data we can significantly constrain the range 
of validity of the dipole picture. 

\end{abstract}

\maketitle

\section{Introduction}
\label{sec-intro}

The HERA collider has provided a wealth of data on deep inelastic 
lepton-proton scattering (DIS), in particular on DIS structure 
functions \cite{Breitweg:2000yn}-\cite{Chekanov:2003yv}. 
Many of these experimental results are analyzed in terms of the dipole 
picture for quasi-real and virtual photon-proton scattering, see for 
instance \cite{Golec-Biernat:1998js}-\cite{Forshaw:2006np}. 
The physical idea underlying the dipole picture 
is that at high energies the quasi-real or virtual photon acts like a 
quark-antiquark dipole \cite{Nikolaev:1990ja,Nikolaev:et,Mueller:1993rr}. 
The $\gamma^* p$ collision is viewed as a two step process.
First the $\gamma^*$ dissociates into a $q\bar q$ dipole, the distribution 
of dipole sizes being described by the photon wave function.
Then the dipole scatters on the proton. 
The latter step is considered to be a purely hadronic reaction. 
The roots of this picture can be traced back to 
\cite{Gribov:1968gs,Ioffe:1969kf}. For a review see \cite{Donnachie:en}. 
In two articles \cite{Ewerz:2004vf,Ewerz:2006vd} the theoretical 
foundations of the standard dipole picture were examined and the 
assumptions were spelled out which are necessary in order to arrive at it. 
Bounds on ratios of DIS structure functions were derived in 
\cite{Ewerz:2006an} which must be respected if the standard dipole 
picture holds. These bounds turn out to be relevant for the important 
question of the range of applicability of the dipole picture. It was in 
fact shown in \cite{Ewerz:2006an} that the comparison of measured 
data on structure functions with the bounds from the dipole picture 
can be used to restrict its range of validity. 

In the present article we discuss further bounds for DIS observables 
which follow from the standard dipole picture.
We shall, in particular, give correlated bounds for $F_L/F_2$ 
versus $F_2^{(c)}/F_2$. Here $F_2,F_2^{(c)}$ and $F_L$ are the 
total $F_2$ structure function, its charm part and the longitudinal 
structure function, respectively. Furthermore, we shall consider 
the structure function $F_2(W,Q^2)$ at the same $\gamma^* p$ 
c.m.\ energy $W$ but at three different values of the photon 
virtuality $Q^2$, and we shall derive correlated bounds on ratios 
of these three values of $F_2$. 
We show that by comparing the bounds found here with measured data 
one can further restrict the kinematical range of applicability of the 
dipole picture. 

Our paper is organized as follows. In section \ref{sec-dipole} we 
fix our notation and recall the main results from 
\cite{Ewerz:2004vf,Ewerz:2006vd} which are relevant here. 
In section \ref{sec-flc2} we derive the correlated bounds on the total, 
longitudinal and charm structure functions. In section \ref{sec-f2q2} 
we discuss the correlated bounds for structure functions at the 
same $\gamma^* p$ c.m.\ energy and three different values of 
$Q^2$. Section \ref{sec-concl} contains our 
conclusions. In appendix \ref{app-hand} we illustrate how our 
bounds are affected at large $Q^2$ by the choice of flux factor in the 
definition of the $\gamma^* p$ cross sections for finite Bjorken-$x$. 
In appendix \ref{app-convex} we explain some mathematical notions and
techniques used in the derivation of our bounds.

\section{The Standard Dipole Picture}
\label{sec-dipole}

We consider deep inelastic lepton-proton scattering
\begin{equation}
\label{processep}
l(k) + p(p) \rightarrow l(k') + X(p')
\end{equation}
where $l=e^-,e^+$. In standard kinematics (see for 
instance \cite{Nachtmann:1990ta}) we have
\begin{equation}
\label{defkin}
\begin{split}
s &= (p + k)^2\,,\\
q &= k - k' = p' - p\,,\\
Q^2 &= -q^2\,,\\
\nu &= p q / m_p\,,\\
W^2 &= (p + q)^2 = 2 m_p \nu - Q^2 + m_p^2\,,\\
y &= \frac{p q}{p k} = \frac{ 2 m_p \nu }{ s - m_p^2 }\,,\\
x &= \frac{ Q^2 }{ 2 m_p \nu }\,.
\end{split}
\end{equation}
We consider moderate $Q^2$,
\begin{equation}
\label{q2vals}
0 \leq Q^2 \lessapprox 10^3 \text{~GeV}^2 \,,
\end{equation}
such that only photon exchange has to be taken into account. 
That is, we are interested in the reaction
\begin{equation}
\label{processap}
\gamma^*(q) + p(p) \rightarrow X(p')\,,
\end{equation}
where the proton is supposed to be unpolarized and a sum over 
all final states $X$ is performed. The total cross section for 
\eqref{processap} is encoded in the hadronic tensor 
\begin{alignat}{2}
\label{hadtens}
&W^{\mu\nu}(p,q) 
= -W_1(\nu,Q^2)\left(g^{\mu\nu}-\frac{q^\mu q^\nu}{q^2}\right) \\
&\;\quad
        +\frac{1}{m^2_p}W_2(\nu,Q^2)\left(p^\mu-\frac{(pq)q^\mu}{q^2}\right)
        \left(p^\nu-\frac{(pq)q^\nu}{q^2}\right) 
\nonumber 
\end{alignat}
with the usual invariant functions $W_{1,2}$.

In order to define the cross sections for longitudinally and transversely 
polarized virtual photons in \eqref{processap} we work in the 
proton rest system, supposing 
\begin{equation}
\label{qprest}
\left( q^\mu \right) =
        \begin{pmatrix} q^0 \\ 0 \\ 0 \\ \abs{\nvec{q}} \end{pmatrix} \,,
\end{equation}
and define the following photon polarization vectors: 
\begin{align}
\label{defepspm}
\left( \varepsilon_\pm^\nu \right) &=
\mp \frac{1}{\sqrt{2}}
\begin{pmatrix} 0 \\ 1 \\ \pm i \\ 0 \end{pmatrix}
\,,\\
\label{defepslong}
\left( \varepsilon_L^\nu \right) &=
\frac{1}{Q}
\begin{pmatrix} \abs{\nvec{q}} \\ 0 \\ 0 \\ q^0 \end{pmatrix}
\,,\\
\label{correctepslong}
\left( \varepsilon_L^{\prime\nu} \right) &=
\left( \varepsilon_L^\nu \right) - \frac{\left(q^\nu\right)}{Q}
=
\frac{1}{Q}
\begin{pmatrix} \abs{\nvec{q}} - q^0 \\ 0 \\ 0 \\ q^0 - \abs{\nvec{q}}
\end{pmatrix}
\,.
\end{align}
With Hand's convention \cite{Hand:1963bb} the $\gamma^* p$ 
cross sections for transverse or longitudinal 
$\gamma^*$ polarization are
\begin{align}
\label{defsigmat}
\sigmatot_T(W,Q^2)
&=
\frac{2\pi m_p}{W^2-m^2_p} \,
\varepsilon^{\mu *}_+e^2W_{\mu\nu}~\varepsilon^\nu_+
\nonumber\\
&=
\frac{2\pi m_p}{W^2-m^2_p}\,\varepsilon^{\mu *}_- 
e^2W_{\mu\nu}~\varepsilon^\nu_-
\nonumber\\
&=
\frac{2\pi m_p}{W^2-m^2_p} \, e^2W_1(\nu,Q^2)
\,,\\
\label{defsigmal}
\sigmatot_L(W,Q^2)
&=
\frac{2\pi m_p}{W^2-m^2_p} \,
\varepsilon'^{\mu *}_L e^2W_{\mu\nu} \, \varepsilon'^\nu_L
\nonumber\\
&=
\frac{2\pi m_p}{W^2-m^2_p}\,
\varepsilon^{\mu*}_L e^2W_{\mu\nu} \, \varepsilon^\nu_L
\nonumber\\
&=
\frac{2\pi m_p}{W^2-m^2_p}\,
\bigg[ e^2 W_2(\nu,Q^2) \,
\frac{\nu^2+Q^2}{Q^2}
\nonumber \\
&
\quad\quad
-e^2 W_1(\nu,Q^2)\bigg] \,.
\end{align}
Note that due to gauge invariance the hadronic tensor 
$W^{\mu\nu}$ \eqref{hadtens} satisfies
\begin{equation}
\label{hadtensginv}
\begin{split}
q_\mu\,  W^{\mu\nu}(p,q) &= 0 \,,\\
W^{\mu\nu}(p,q)\, q_\nu &= 0 \,.
\end{split}
\end{equation}
Thus in the definition of $\sigmatot_L$ \eqref{defsigmal} it is 
irrelevant whether we choose the $\gamma^*$ polarization vector as 
$\varepsilon^\nu_L$ \eqref{defepslong} or 
$\varepsilon'^\nu_L$ \eqref{correctepslong}. 
However, as shown in \cite{Ewerz:2006vd}, 
in applications of the dipole model it is {\em essential} to use 
$\varepsilon'^\nu_L$ and {\em not} $\varepsilon^\nu_L$, 
in particular when one calculates the photon wave function 
from the Feynman rule for an incoming photon splitting into 
outgoing on-shell quark and antiquark. In that case the photon 
polarization vector has to be chosen such that its components 
remain finite in the high energy limit, as is true for $\varepsilon'_L$ 
but not for $\varepsilon_L$. 

The standard structure function $F_2$ is defined as
\begin{alignat}{2}
&F_2(W, Q^2) = \nu\, W_2(\nu,Q^2)
\nonumber \\
\label{f2diphand}
&\;= \frac{Q^2}{4 \pi^2 \alpha_{\rm em}}
        \left[ \sigmatot_T(W,Q^2) + \sigmatot_L(W,Q^2) \right]
\nonumber\\
&\;\quad\times
        \left\{ 1 +
                \frac{ Q^2 \left( W^2 + Q^2 + 3 m_p^2 \right) }
                     { \left( W^2 - m_p^2 \right)\left( W^2 + Q^2 - m_p^2 \right) }
        \right\}^{-1}
\nonumber \\
&\;= \frac{Q^2}{4 \pi^2 \alpha_{\rm em}}
        \left[ \sigmatot_T(W,Q^2) + \sigmatot_L(W,Q^2) \right] (1-x)
\nonumber\\
&\;\quad + \mathcal{O}(m_p^2/W^2)
        \,.
\end{alignat}
In the high energy limit, $W \gg Q, m_p$, 
this simplifies to the commonly used form 
\begin{equation}
\label{f2dipsimple}
F_2(W, Q^2) =
        \frac{Q^2}{4 \pi^2 \alpha_{\rm em}}
                \left[ \sigmatot_T(W,Q^2) + \sigmatot_L(W,Q^2) \right]
\end{equation}
up to terms of order $\mathcal{O}(Q^2/W^2)$. 
Similarly, we use for the standard longitudinal structure function 
\begin{equation}
\label{fldip}
F_L(W,Q^2)
= \frac{Q^2}{4 \pi^2 \alpha_{\rm em}}
        \sigmatot_L(W,Q^2)
\,.
\end{equation}

In the following we shall use the relation \eqref{f2dipsimple} 
valid in the high energy limit. In appendix \ref{app-hand} 
we shall discuss how our results are modified for finite 
Bjorken-$x$ if we use the exact formula \eqref{f2diphand} 
instead of \eqref{f2dipsimple}. 
We note that one could also consider \eqref{f2dipsimple} 
as the defining equation for $\sigma_T$ and $\sigma_L$. 
This would correspond to a different choice of flux factor 
for the virtual photons as compared to \cite{Hand:1963bb}. 
The considerations of section 6 of \cite{Ewerz:2006vd} 
show, however, that Hand's convention \cite{Hand:1963bb} 
is the natural one for the dipole picture; see especially 
(121)-(128) of \cite{Ewerz:2006vd}. 

In \cite{Ewerz:2004vf,Ewerz:2006vd} nonperturbative methods were 
employed in order to work towards a foundation of the dipole model 
for quasi-real and virtual photon induced reactions at high energies. 
The result for 
$W^{\mu\nu}$ (2.5) obtained there is shown diagrammatically 
in Fig.~\ref{fig-classes}. In the high energy limit, 
$q^0\to\infty$, taken in the proton rest frame, we find a factorization 
into photon wave function and dipole-proton scattering parts. 
The wave function parts contain the renormalized $\gamma q \bar{q}$ 
vertex function plus a rescattering term. The dipole-proton scattering 
is built from diagrams of type (a) where the quark lines go through 
from right to left and type (b) where the quark lines do not go through. 
To get from there to the standard dipole picture requires to make 
a number of {\sl assumptions} and approximations 
as listed in \cite{Ewerz:2006vd}: 
\begin{figure*}
\includegraphics[width=\textwidth]{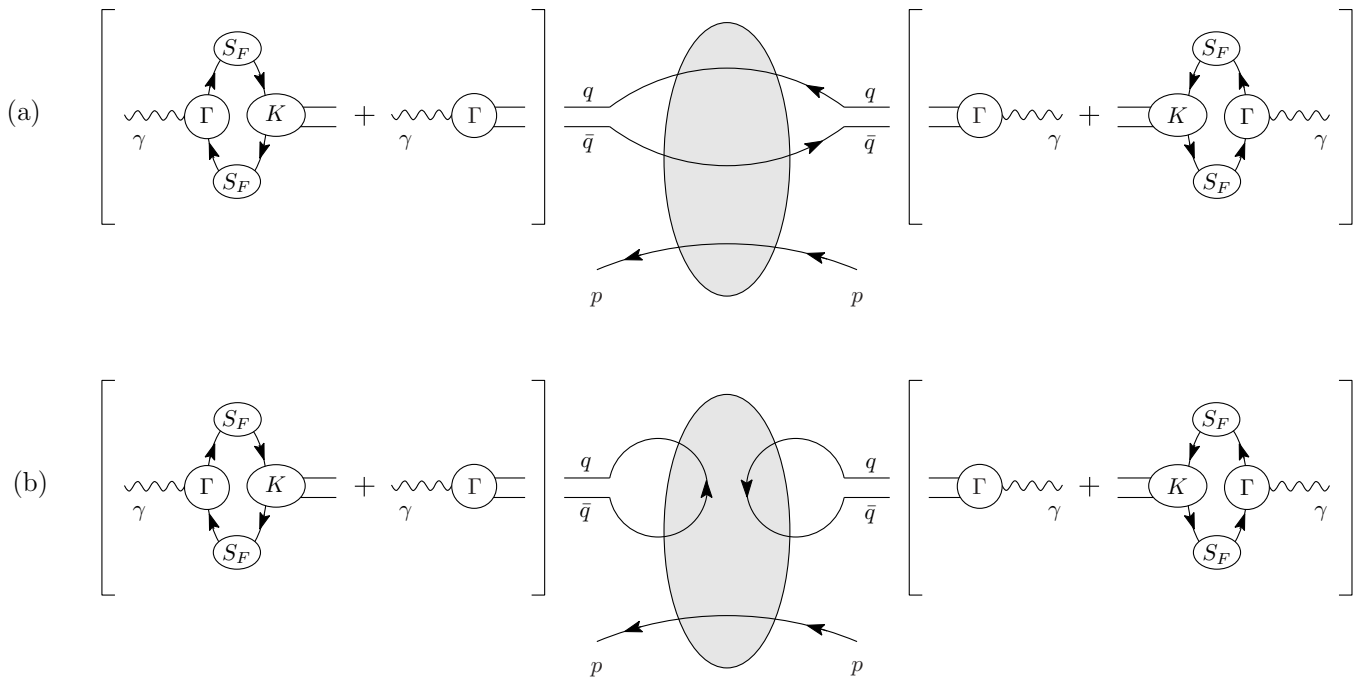}
\caption{
\label{fig-classes}
Quark skeleton diagrams for the photon-proton scattering 
cross section in the high energy limit. 
The shaded area indicates a functional integration over gluon field 
configurations, 
and $\Gamma$, $K$, $S_F$ are the renormalized $\gamma q \bar{q}$ 
vertex, the renormalized kernel for $q'\bar{q}'$ to $q\bar{q}$ scattering, 
and the renormalized quark propagator, respectively.
The diagrams are to be read from right to left. 
}
\end{figure*}

\begin{itemize}
\item [(i)]
Quarks of flavor $q$ have a mass shell $m_q$ and can be 
considered as asymptotic states. 
\item [(ii)]
The rescattering terms are dropped and the $\gamma q \bar{q}$ vertex 
functions are replaced by the lowest order terms in perturbation theory. 
\item [(iii)]
The $T$-matrix element for the dipole-proton scattering is diagonal 
in the quark flavor $q$, in $\alpha$ and in $\nvec{r}$. Here $\alpha$ is 
the longitudinal momentum fraction of the photon carried by the 
quark, and $\nvec{r}$ is the two-dimensional vector from the antiquark
to the quark in transverse position space. 
Further, the $T$-matrix element is proportional 
to the unit matrix in the space of spin orientations of the quark and 
antiquark in the dipole. 
\item [(iv)]
In the $T$-matrix element for the dipole-proton scattering  only the 
contribution of type (a) is kept while that of type (b) is neglected, see 
Fig.~\ref{fig-classes}.
\item [(v)]
The proton spin averaged reduced matrix element for a given 
quark flavor $q$ depends only on the dipole size 
$r\equiv\sqrt{\nvec{r}^2}$ and on $W^2=(p+q)^2$. 
\end{itemize}

With these assumptions we arrive indeed at the standard formulae of 
the dipole picture used extensively in the literature. The squared and 
spin-summed photon wave functions for quark flavor $q$ can be 
calculated in leading order in $\alpha_{\rm em}$ resulting in 
\begin{alignat}{2}
\label{sumpsitdens}
\lefteqn{
        \sum_{\lambda, \lambda'} \left| 
        \psi_{\gamma, \lambda \lambda'}^{(q) \mu} 
          (\alpha, \nvec{r},Q) \,\varepsilon_{+\mu}\right|^2 
}\qquad&&&
\nonumber \\
&&=
        \frac{N_c}{2 \pi^2} \, \alpha_{\rm em} Q_q^2 
        \big\{ & \left[ \alpha^2 + (1-\alpha)^2 \right] 
        \epsilon_q^2 [K_1(\epsilon_q r) ]^2 \nonumber\\
        &&&+ m_q^2 [K_0(\epsilon_q r) ]^2 
        \big\} \,,
\\
\label{sumpsildens}
\lefteqn{
        \sum_{\lambda, \lambda'} \left| 
        \psi_{\gamma, \lambda \lambda'}^{(q) \mu}(\alpha, \nvec{r},Q) 
         \, \varepsilon'_{L \mu}\right|^2 
}\qquad&&&
\nonumber \\
&&=
        \frac{2 N_c}{\pi^2} \, \alpha_{\rm em} Q_q^2
        & Q^2 [\alpha (1-\alpha)]^2 [K_0(\epsilon_q r) ]^2 
\end{alignat}
for transversely and longitudinally polarized photons, respectively. 
Here $N_c = 3$ is the number of colors,
$\epsilon_q = \sqrt{\alpha (1-\alpha) Q^2 +m_q^2}$, 
$Q_q$ denotes the quark charges in units of the proton charge,
and $K_{0,1}$ are modified Bessel functions. 
Upon integration over $\alpha$ we obtain from the above expressions 
the photon densities as functions of the dipole size $r$ and of $Q^2$, 
\begin{align}
\label{denst}
w^{(q)}_T( r,Q^2) &=
\sum_{\lambda,\lambda'}\int^1_0 \ud\alpha \,
\left|
\psi^{(q)\mu}_{\gamma,\lambda\lambda'}(\alpha,\nvec{r},Q)
\,\varepsilon_{+\mu} \right|^2 \,,
\\
\label{densl}
w^{(q)}_L( r,Q^2) &=
\sum_{\lambda,\lambda'}\int^1_0 \ud\alpha\,
\left| 
\psi^{(q)\mu}_{\gamma,\lambda\lambda'}(\alpha,\nvec{r},Q)
\,\varepsilon'_{L\mu} \right|^2 \,.
\end{align}
The expressions for the $\gamma^{(*)} p$ total cross sections in the 
standard dipole picture are 
\begin{align}
\label{sigmatdip}
\sigmatot_T(W,Q^2)&=
\sum_q\int \ud^2 r \,
w^{(q)}_T( r,Q^2)\,
\sigmadip^{(q)}( r,W) \,,
\\
\label{sigmaldip}
\sigmatot_L(W,Q^2) &=
\sum_q\int \ud^2 r \,
w^{(q)}_L( r,Q^2)\,
\sigmadip^{(q)}( r,W) \,.
\end{align}
Here $\sigmadip^{(q)}( r,W)$ is the cross section for the 
scattering of a dipole of flavor $q$ and size $r$ on a proton 
for a dipole-proton c.m.\ energy $W$. 
Presently, the dipole-proton cross section cannot be calculated 
from first principles and one therefore uses models with parameters 
obtained by fitting the available data. 

Note that the correct energy variable of the dipole-proton cross section 
$\sigmadip$ is $W$ and not Bjorken-$x$. The latter would imply 
a dependence on the photon virtuality $Q^2$. It was argued in 
\cite{Ewerz:2006vd} that using $x$ instead of $W$ requires additional 
assumptions which are difficult to assess quantitatively and 
which go beyond those listed above. 
Nevertheless, the energy variable $x$ -- and hence a $Q^2$-dependence -- 
is frequently used in popular models for the dipole cross section (and 
sometimes other dependencies on $Q^2$ are introduced). 

In the following we shall also consider $F_2^{(c)}$, that is, the 
structure function for $\gamma^*p$ scattering with production 
of charm particles. In the dipole model the cross sections for charm 
production are obtained as in \eqref{sigmatdip}, \eqref{sigmaldip} but 
without the summation over quark flavor $q$, 
\begin{equation}
\label{sigmacdip}
\sigmatot^{(c)}_{T,L}(W,Q^2) = \int \ud^2 r\, w^{(c)}_{T,L}( r,Q^2)\,
\sigmadip^{(c)}( r,W)\,.
\end{equation}
We set as in \eqref{f2dipsimple} for the charm structure function 
\begin{equation}
\label{f2cdip}
F_2^{(c)} = \frac{Q^2}{4 \pi^2 \alpha_{\rm em}} \left[
        \sigmatot^{(c)}_T(W,Q^2) + \sigmatot^{(c)}_L(W,Q^2)\right]
\,.
\end{equation}
Note that in doing so we make the assumption that all charm quarks 
produced originate directly from the initial $\gamma^*$. That is, we 
neglect associated charm-anticharm production in reactions initiated 
by other quark flavors coupling directly to the photon. 

In the following sections we shall use \eqref{sigmatdip}-\eqref{sigmacdip} 
to derive bounds for ratios of structure functions. These bounds will rest 
on the explicit forms of the photon densities $w^{(q)}_{T,L}$ 
\eqref{denst}, \eqref{densl} and on the non-negativity of the 
dipole-proton cross sections 
\begin{equation}
\label{sigmadippos}
\sigmadip^{(q)}(r,W) \geq 0 \,.
\end{equation}
The bounds derived in Sect.~\ref{sec-flc2} remain unchanged if we 
assume that the dipole cross sections $\sigmadip^{(q)}$ are functions 
of $r$ and Bjorken-$x$ instead of $r$ and $W$. The bounds derived in 
Sect.~\ref{sec-f2q2}, on the other hand, depend crucially on the 
functional dependence indicated in \eqref{sigmadippos}.

\boldmath
\section{Bounds for $F_L/F_2$ and $F_2^{(c)}/F_2$}
\unboldmath
\label{sec-flc2}

In this section we consider the structure functions $F_L$, 
$F_2^{(c)}$, and $F_2$ at fixed values of $Q^2$ and $W$. 
Arranging them into a three-vector gives 
according to the dipole formula 
\begin{equation}
        \label{flc2vec}
\mbox{\small $ 
       \begin{pmatrix}
                 F_L(W,Q^2) \\
                F_2^{(c)}(W,Q^2) \\
                 F_2(W,Q^2) 
        \end{pmatrix}
$}
        =
        \sum_{q} \int \ud^2 r\,
        \frac{\hat{\sigmatot}^{(q)}(r,W)}{4\pi^2\alpha_{\text{em}}}
\mbox{\small $ 
        \begin{pmatrix}
                \phantom{\delta_{q,c}}\, f^{(q)}_L(r,Q^2)\\
                \delta_{q,c}\, f^{(c)}(r,Q^2)\\
                \phantom{\delta_{q,c}}\, f^{(q)}(r,Q^2)
        \end{pmatrix}
$}
\,,
\end{equation}
with 
\begin{align}
\label{f2fdef}
f^{(q)}(r,Q^2) &= Q^2 \left[w_T^{(q)}(r,Q^2)+w_L^{(q)}(r,Q^2)\right]\,,
\\
\label{f2fdefII}
f^{(q)}_L(r,Q^2) &= Q^2 \,w_L^{(q)}(r,Q^2)\,.
\end{align}
Note that the second entry in the vector in \eqref{flc2vec} 
receives a contribution only 
from the charm quark, as indicated by the Kronecker delta symbol. 
In the following we will make use of a geometrical interpretation 
of \eqref{flc2vec} in order to obtain correlated bounds on the structure 
functions involved here. Due to the Kronecker 
symbol the case at hand is somewhat special, which might make 
the geometrical interpretation slightly more difficult to conceive. 
An illustration of the general argument is given in Fig.\ 
\ref{fig-vecs4f2dratios} in section \ref{sec-f2q2} below where 
we discuss similar three-vectors of structure functions, but there 
without the occurrence of a Kronecker symbol. 

We recall that the dipole cross sections $\hat{\sigma}^{(q)}$ are non-negative. 
Thus the r.h.s.\ of \eqref{flc2vec} is a sum and an integral over
three-vectors multiplied by non-negative weights, or, in other words, 
a special linear superposition of the three-vectors appearing under the integral.
We want to find the set of all possible linear superpositions of this kind with 
non-negative coefficients.
This is called a moment problem. In appendix \ref{app-convex} we discuss
the necessary mathematical tools to solve this problem.
We give there the precise definitions of the key concepts convex set,
convex hull and convex cone. We also give the detailed solution of the
moment problem for the case of three $F_2^{(q)}$ structure functions
as discussed below in section \ref{sec-f2q2}.
The solution of the moment problem in this section runs along the same lines.
The analogue of the result \eqref{B.35b} reads here as follows.
The set of all vectors allowing a representation \eqref{flc2vec} is given by a
convex cone. 
Any vector within this cone can be written as a non-negative multiple of an 
element within the closed convex hull (denoted by $\cohullcl$)
of the three-vectors appearing in the r.h.s.\ of \eqref{flc2vec}. 
Therefore we have 
\begin{widetext}
\begin{equation}
\label{flc2convexcond}
        \begin{pmatrix}
                F_L (W,Q^2) \\
                F_2^{(c)}(W,Q^2) \\
                F_2(W,Q^2)
        \end{pmatrix}
= \lambda \nvec{u}(Q^2)\,,
\quad
\lambda \ge 0 \,,
\quad
  \nvec{u}(Q^2)
  \in \cohullcl \left\{\left.
  \begin{pmatrix}
                \phantom{\delta_{q,c}}\, f^{(q)}_L(r,Q^2)\\
                \delta_{q,c}\, f^{(c)}(r,Q^2)\\
                \phantom{\delta_{q,c}}\, f^{(q)}(r,Q^2)
  \end{pmatrix}
    \,\right\vert\,r \in \realnums^+ , q=u,d,\ldots \right\}
\,.
\end{equation}
\end{widetext}
Note that the three-vectors from which the convex hull is 
constructed involve only the functions $f^{(q)}(r,Q^2)$ and 
$f^{(q)}_L(r,Q^2)$ which are for any given $Q^2$ explicitly 
known for all $r$, see \eqref{sumpsitdens}-\eqref{densl}. 
Hence it is also straightforward to compute their convex hull. 
We further point out that these vectors are independent of 
the energy $W$, and that the condition \eqref{flc2convexcond} 
does not involve any model assumption about the dipole 
cross section $\hat{\sigma}^{(q)}$. 

We can now use the condition \eqref{flc2convexcond} to derive 
bounds on ratios of $F_L$, $F_2^{(c)}$, and $F_2$. These 
bounds originate 
only from the photon wave functions. They will be valid for 
any dipole cross section $\sigmadip^{(q)}$, and will 
be independent of the energy $W$. Clearly, the bounds will 
vary  with the photon virtuality $Q^2$, since $Q^2$ explicitly 
enters the vectors in \eqref{flc2convexcond} via the photon 
wave function. 

We first notice that the condition \eqref{flc2convexcond} 
constrains only the {\sl directions} of the three-vectors 
involved, while their normalization is irrelevant for that 
condition. We can therefore normalize the vector 
composed of the three structure functions such that its 
third component equals one, that is, we consider the 
vector $(F_L/F_2, F_2^{(c)}/F_2, 1)^{\rm T}$ instead 
of $(F_L, F_2^{(c)}, F_2)^{\rm T}$. That normalization 
does not change the direction of the vector, and hence also 
the so normalized vector fulfills the condition 
\eqref{flc2convexcond}. Similarly, we can also normalize 
the set of vectors of which the closed convex hull is formed such 
that its third component equals one, hence considering 
$(f^{(q)}_L/f^{(q)}, \delta_{q,c}f^{(c)}/f^{(q)},1)^{\rm T} = 
(f^{(q)}_L/f^{(q)}, \delta_{q,c},1)^{\rm T}$ 
in place of $(f^{(q)}_L, \delta_{q,c}f^{(c)}, f^{(q)})^{\rm T}$. 
Again, that does not affect the direction of the vectors, 
and the condition \eqref{flc2convexcond} immediately 
applies with this replacement. The condition with both 
vectors normalized in this way contains only vectors the third 
component of which equals one, and for this case the 
only possible choice for the factor $\lambda$ is $\lambda=1$. 
We can then eliminate the trivial third component by 
projecting onto the 1-2-plane and obtain from 
\eqref{flc2convexcond} the simpler condition 
\begin{widetext}
\begin{equation}
\label{flc2corbnds}
\begin{pmatrix} F_L(W,Q^2)/F_2(W,Q^2) \\
                F_2^{(c)}(W,Q^2)/F_2(W,Q^2) \end{pmatrix} \in 
\cohullcl \left\{\left.
  \begin{pmatrix}
   f^{(q)}_L(r,Q^2)/f^{(q)}(r,Q^2)\\
   \delta_{q,c}\end{pmatrix}
    \,\right\vert\, r \in \realnums^+ , q=u,d,\ldots \right\}
\,,
\end{equation}
\end{widetext}
which is in fact equivalent to the original condition 
\eqref{flc2convexcond} for the realistic case that $F_2$ and 
$f^{(q)}$ for $r \in \realnums^+$ are strictly positive.
For a rigorous derivation of bounds on ratio vectors as in
\eqref{flc2corbnds} see appendix \ref{app-convex}, where the analogous
case of three $F_2$ structure functions is discussed in detail
(cf. \eqref{B.38}).

The first bound that we want to discuss here is now obtained 
from \eqref{flc2corbnds} by projecting onto the 1-axis. 
This immediately gives 
\begin{equation}
\label{fl2uncbnds}
 \inf_{r, q}  \frac{f^{(q)}_L(r,Q^2)}{f^{(q)}(r,Q^2)} \leq
              \frac{ F_L(W,Q^2) }{ F_2(W,Q^2) } \leq
 \sup_{r, q}  \frac{f^{(q)}_L(r,Q^2)}{f^{(q)}(r,Q^2)}
\,,
\end{equation}
where $\inf$ and $\sup$ denote the infimum and supremum, 
respectively \footnote{We recall that for a given subset $S$ of 
$\realnums$ the infimum of the set $S$ is the greatest number 
less than or equal to each element of $S$. Similarly, the 
supremum is the smallest number that is greater than or equal 
to each element of $S$. For a compact set $S$ the minimum (maximum) 
coincides with the infimum (supremum).}. 
Note that these lower and upper bounds on $F_L/F_2$ are given 
only in terms of the photon wave function. 
It is therefore straightforward to analyze the bounds \eqref{fl2uncbnds} 
numerically. 

\begin{figure}
\includegraphics[width=\columnwidth]{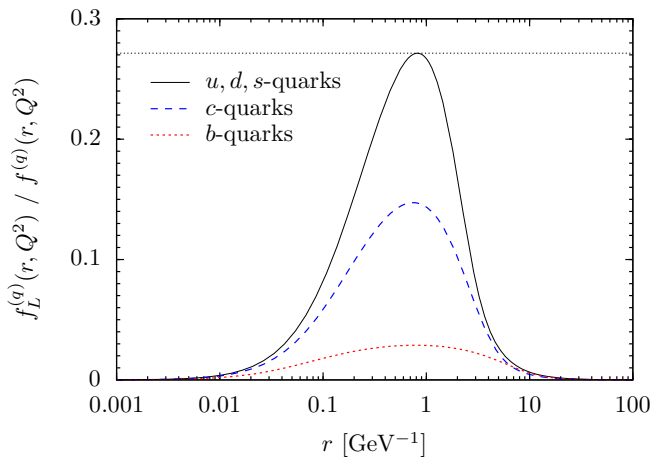}
\caption{
\label{fig-ratOfLgRFLF2}
The ratio $f^{(q)}_L(r,Q^2)/f^{(q)}(r,Q^2)$ as a function of $r$ 
for different quark flavors. The photon virtuality is chosen to be 
$Q^2=10\text{~GeV}^2$. 
The absolute maximum value (dotted line) of all curves
provides an upper bound on $F_L(W,Q^2)/F_2(W,Q^2)$, 
see \eqref{fl2uncbnds} and \eqref{numflf2bound}. 
}
\end{figure}

In Fig.~\ref{fig-ratOfLgRFLF2} we plot the ratio 
$f^{(q)}_L(r,Q^2)/f^{(q)}(r,Q^2)$ as a function of $r$ for different 
quark flavors, choosing as an example $Q^2 = 10\,\mbox{GeV}^2$. 
Here and in the following we use vanishing masses for the light ($u,d,s$) quarks,
$m_c=1.3\text{~GeV}$ for the charm quark and $m_b=4.6\text{~GeV}$ for the 
bottom quark. We find that the lower bound in \eqref{fl2uncbnds} is 
trivial, $F_L/F_2 \ge 0$. 
The upper bound, on the other hand, is nontrivial, and we find that the 
maximal value of $f^{(q)}_L(r,Q^2)/f^{(q)}(r,Q^2)$ is obtained for light 
quarks, as can be seen in Fig.~\ref{fig-ratOfLgRFLF2} where this maximum 
is drawn as a dotted horizontal line. It turns out that this upper bound 
is independent of $Q^2$ and numerically leads to
\begin{equation}
\label{numflf2bound}
 \frac{F_L(W,Q^2)}{F_2(W,Q^2)} \,\le\, 0.27139 \,.
\end{equation}

\begin{figure}
\includegraphics[width=\columnwidth]{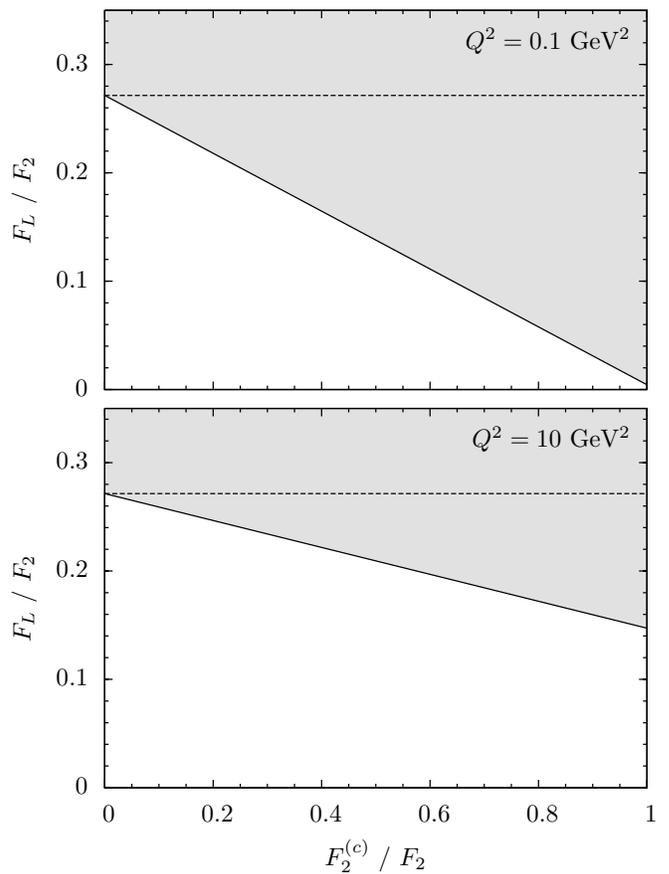}
\caption{
\label{fig-f2corbnds}
Correlated bounds on $F_L/F_2$ and $F_2^{(c)}/F_2$. 
Only the white (unshaded) area is allowed within the dipole picture.
The weaker bound \eqref{fl2uncbnds} is shown as a dashed 
line, while the stronger bound implied by \eqref{flc2corbnds} is 
shown as a solid line. 
}
\end{figure}

A stronger bound can be obtained by considering the 
correlation of the ratios $F_L/F_2$ and $F_2^{(c)}/F_2$, 
that is, by taking into account both components of the 
constraint on the vectors in \eqref{flc2corbnds}. 
In this case the bound on the ratio $F_L/F_2$ will depend 
on the value of $F_2^{(c)}/F_2$ or vice versa. 
By computing the closed convex hull in \eqref{flc2corbnds} from 
the ratios $f^{(q)}_L(W,Q^2)/f^{(q)}(W,Q^2)$ we obtain the 
correlated bounds shown in Fig.\ \ref{fig-f2corbnds} 
for the two values $Q^2 = 0.1\,\mbox{GeV}^2$ and 
$Q^2 = 10\,\mbox{GeV}^2$. 
The unshaded area in the two plots is the allowed region 
in the dipole picture. The Figure is drawn for the whole 
range of $F_2^{(c)}/F_2$ between zero and one to make the 
origin of the bounds clear. Realistic values of $F_2^{(c)}/F_2$ 
can only range from zero to at most about 0.4. 
The allowed area in Fig.\ \ref{fig-f2corbnds} 
is bounded by a straight line. This particular shape emerges 
due to the fact that the second component of the vectors in 
\eqref{flc2corbnds} receives a contribution only from the charm 
quark. Due to the corresponding Kronecker symbol the upper 
bound on $F_L/F_2$ at the (unphysical) point $F_2^{(c)}/F_2 =1$ 
is given by the maximum of $f^{(c)}_L(r,Q^2)/f^{(c)}(r,Q^2)$ over 
all $r$ for the $Q^2$ under consideration. The value of this maximum 
for the case $Q^2=10\,\mbox{GeV}^2$ can be read off from the 
charm quark curve in Fig.\ \ref{fig-ratOfLgRFLF2}. For $Q^2$-values 
well below the charm mass, like for example $Q^2=0.1\,\mbox{GeV}^2$, 
the analogous function practically vanishes and the resulting 
upper bound on $F_L/F_2$ at $F_2^{(c)}/F_2 =1$ is practically zero. 
The fact that the unphysical point $F_2^{(c)}/F_2 =1$ is relevant 
for the determination of the correlated bounds on $F_L/F_2$ and 
$F_2^{(c)}/F_2$  in the physical region of these ratios should 
not cause any worries here. It is just the consequence of not making 
any assumptions about the flavor dependence of the dipole cross 
sections $\sigmadip^{(q)}$, except their being non-negative. 
This general case includes for example the unphysical case that 
all dipole cross sections but the one for the charm quark would vanish, 
which would give rise to $F_2^{(c)}/F_2 =1$. By making further 
assumptions about the dipole cross sections it should be possible to 
derive more stringent bounds -- but at the expense of introducing 
a dependence on those assumptions. 
In the present paper, however, it is our aim 
to study bounds on ratios of structure functions from the dipole 
picture which {\sl do not} depend on any further assumptions 
on the dipole cross section. 

Future measurements of the structure functions $F_L$, 
$F_2^{(c)}$and $F_2$ at identical values of $Q^2$ and $W$ might 
in combination with our bounds be able to constrain the range of 
validity of the dipole picture. 

Closing this section we would like to point out again that the geometric 
argument and its implications discussed in this section remain unchanged 
if the dipole cross section is chosen to depend on $x$ instead of $W$. 

\boldmath
\section{Bounds on Ratios of $F_2$ at Different Values of $Q^2$}
\unboldmath
\label{sec-f2q2}

In this section we use the dipole picture to derive bounds on 
ratios of the structure function $F_2$ taken at the same 
$W$ but at different values of $Q^2$. The results found here 
crucially depend on choosing the functional 
dependence of the dipole cross section such that its 
arguments are $r$ and $W$. 
In particular, the dipole cross section $\sigmadip^{(q)}(r,W)$ 
is assumed to be independent of $Q^2$, see the corresponding 
discussion in section \ref{sec-dipole}. 

We consider the structure function $F_2$ at three different values 
of $Q^2$ but at the same $W$. Similarly to the previous section we 
arrange them into a three-vector, and evaluate it according to 
the dipole formula, 
\begin{equation}
\label{f2threeQ2vec}
\begin{pmatrix} F_2(W,Q_1^2) \\
                F_2(W,Q_2^2) \\
                F_2(W,Q_3^2) \end{pmatrix}=
\sum_{q} \int \ud^2 r\,
\frac{\hat{\sigmatot}^{(q)}(r,W)}{4\pi^2\alpha_{\text{em}}}
\begin{pmatrix}
 f^{(q)}(r,Q_1^2)\\
 f^{(q)}(r,Q_2^2)\\
 f^{(q)}(r,Q_3^2)
\end{pmatrix}\,,
\end{equation}
where the $f^{(q)}(r,Q_i^2)$ are defined in \eqref{f2fdef}. 
We can now derive bounds on ratios of such structure functions 
following the same procedure as in the preceeding section. 
To find all vectors allowing a representation \eqref{f2threeQ2vec}
is again a moment problem. In appendix \ref{app-convex} we discuss
the solution of this problem for the case at hand in a
mathematically rigorous way. A simple argument, leaving out some
subtleties, is as follows.

\begin{figure}[htb]
\includegraphics[width=0.7\columnwidth]{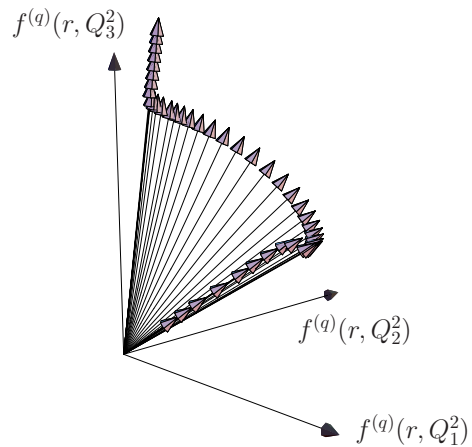}
\caption{
\label{fig-vecs4f2dratios}
The vectors $(f^{(q)}(r,Q_1^2), f^{(q)}(r,Q_2^2), f^{(q)}(r,Q_3^2))^{\rm T}$ 
for different values of $r$, shown here for a massless quark flavor $q$ and 
for one particular choice of the triple $(Q_1^2, Q_2^2, Q_3^2)$. 
}
\end{figure}

The vector on the l.h.s.\ of \eqref{f2threeQ2vec} is a linear 
superposition of the vectors 
$(f^{(q)}(r,Q_1^2),f^{(q)}(r,Q_2^2), f^{(q)}(r,Q_3^2))^{\rm T}$ 
which appear under the integral. 
For a given flavor $q$ and given values of the $Q_i^2$ that 
vector follows a trajectory as $r \in \realnums^+$ is varied. 
Fig.\ \ref{fig-vecs4f2dratios} illustrates a number of vectors 
along such a trajectory for the case of massless quarks and for 
one particular choice of $Q_1^2$, $Q_2^2$, and $Q_3^2$. 
We recall again that the dipole cross sections $\sigmadip^{(q)}$ 
are non-negative. Accordingly, the r.h.s.\ of 
\eqref{f2threeQ2vec} is a linear superposition with non-negative weights 
of the vectors that appear under the integral. Therefore the 
resulting vector on the l.h.s.\ must lie in the closed convex cone 
formed by all possible linear superpositions with non-negative weights 
of those vectors and their boundary. Any vector within such a cone is 
a non-negative multiple of a vector that lies in the closed convex hull
(denoted by $\cohullcl$) of the vectors appearing under the integral
in \eqref{f2threeQ2vec}, see \eqref{B.35b} of appendix~\ref{app-convex}.
Hence we obtain the condition 
\begin{widetext}
\begin{equation}
\label{f2convexcond}
\begin{pmatrix} F_2(W,Q_1^2) \\
                F_2(W,Q_2^2) \\
                F_2(W,Q_3^2) \end{pmatrix} = 
\mu  \nvec{v}(Q_1^2,Q_2^2,Q_3^2)\,,
\quad
\mu \ge 0 \,,
\quad
  \nvec{v}(Q_1^2,Q_2^2,Q_3^2)
  \in \cohullcl \left\{\left.
  \begin{pmatrix}
   f^{(q)}(r,Q_1^2)\\
   f^{(q)}(r,Q_2^2)\\
   f^{(q)}(r,Q_3^2)
  \end{pmatrix}
    \right\vert\,r \in \realnums^+ , q=u,d,\ldots \right\}
\,.
\end{equation}
\end{widetext}
As in the case of \eqref{flc2corbnds} in the previous section this 
condition constrains only the directions of the vectors involved, 
but not their length. This applies both to the vector with 
components $F_2(W,Q_i^2)$ and to the vector with 
components $f^{(q)}(r,Q_i^2)$. Accordingly, we can normalize these 
vectors such that their third component equals one. Doing this 
for both vectors that appear in \eqref{f2convexcond} we obtain an 
equivalent condition. In this condition the only possible value 
for $\mu$ is obviously $\mu =1$. Since the third component 
of the condition is now trivial we discard it by projecting onto 
the 1-2-plane to obtain 
\begin{widetext}
\begin{equation}
\label{f2corbnds}
\begin{pmatrix} F_2(W,Q_1^2)/F_2(W,Q_3^2) \\
                F_2(W,Q_2^2)/F_2(W,Q_3^2) \end{pmatrix} \in
\cohullcl \left\{\left.
  \begin{pmatrix}
   f^{(q)}(r,Q_1^2)/f^{(q)}(r,Q_3^2)\\
   f^{(q)}(r,Q_2^2)/f^{(q)}(r,Q_3^2) \end{pmatrix}
    \,\right\vert\,r \in \realnums^+ , q=u,d,\ldots \right\} \,,
\end{equation}
\end{widetext}
which is fully equivalent to \eqref{f2convexcond} because 
$F_2(W,Q_3^2) $ and $f^{(q)}(r,Q_3^2)$ are strictly positive
for the relevant range of their arguments.
For a rigorous derivation of \eqref{f2corbnds} see
\eqref{B.36}-\eqref{B.38} of appendix \ref{app-convex}.

Let us first consider the two components of the 
condition \eqref{f2corbnds} separately. 
Projecting it onto the 1-axis  and onto the 2-axis 
immediately gives the conditions 
\begin{align}
\label{f2uncbnds13}
 \inf_{r, q}  \frac{f^{(q)}(r,Q_1^2)}{f^{(q)}(r,Q_3^2)} &\leq
              \frac{ F_2(W,Q_1^2) }{ F_2(W,Q_3^2) } \leq
 \sup_{r, q}  \frac{f^{(q)}(r,Q_1^2)}{f^{(q)}(r,Q_3^2)},\\
\label{f2uncbnds23}
 \inf_{r, q}  \frac{f^{(q)}(r,Q_2^2)}{f^{(q)}(r,Q_3^2)} &\leq
              \frac{ F_2(W,Q_2^2) }{ F_2(W,Q_3^2) } \leq
 \sup_{r, q}  \frac{f^{(q)}(r,Q_2^2)}{f^{(q)}(r,Q_3^2)} \,,
\end{align}
respectively. The condition \eqref{f2uncbnds23} goes into 
\eqref{f2uncbnds13} if we replace 
$Q_2^2$ by $Q_1^2$. Thus, for two given values of $Q^2$ 
we actually obtain one condition here which contains an upper and 
and a lower bound. The same result 
was already presented in \cite{Ewerz:2006an}, where 
these bounds were derived in a different way. 

\begin{figure}
\includegraphics[width=\columnwidth]{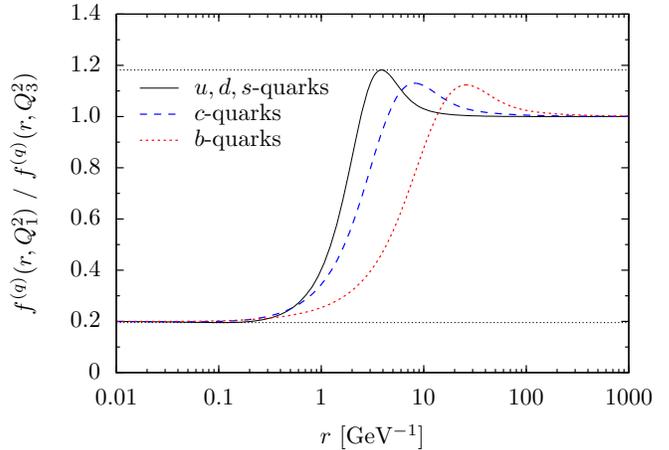}
\caption{
\label{fig-ratOfLgRF2}
The ratio $f^{(q)}(r,Q_1^2)/f^{(q)}(r,Q_3^2)$ as a function of $r$
for different quark masses, here for the choice 
$(Q_1^2,Q_3^2)=(2,10)\text{~GeV}^2$. 
The minimum and maximum values of all curves (shown 
as dotted lines) provide a lower and
upper bound on $F_2(W,Q_1^2)/F_2(W,Q_3^2)$, respectively, 
see \eqref{f2uncbnds13}.
}
\end{figure}

We find it useful to discuss here briefly the bounds \eqref{f2uncbnds13}, 
for a more detailed discussion we refer the reader to \cite{Ewerz:2006an}. 
We first note that the upper and lower bound \eqref{f2uncbnds13} 
depend only on the values of $Q_1^2$ and $Q_3^2$, but do 
not involve the energy $W$. 
Fig.~\ref{fig-ratOfLgRF2} shows the ratio 
$f^{(q)}(r,Q_1^2)/f^{(q)}(r,Q_3^2)$ as a function of $r$ for different 
quark masses along with the resulting bounds on 
$F_2(W,Q_1^2)/F_2(W,Q_3^2)$ 
for a concrete choice of $Q_1^2,Q_3^2$, 
$(Q_1^2,Q_3^2)=(2,10)\,\mbox{GeV}^2$. 

\begin{figure}
\includegraphics[width=\columnwidth]{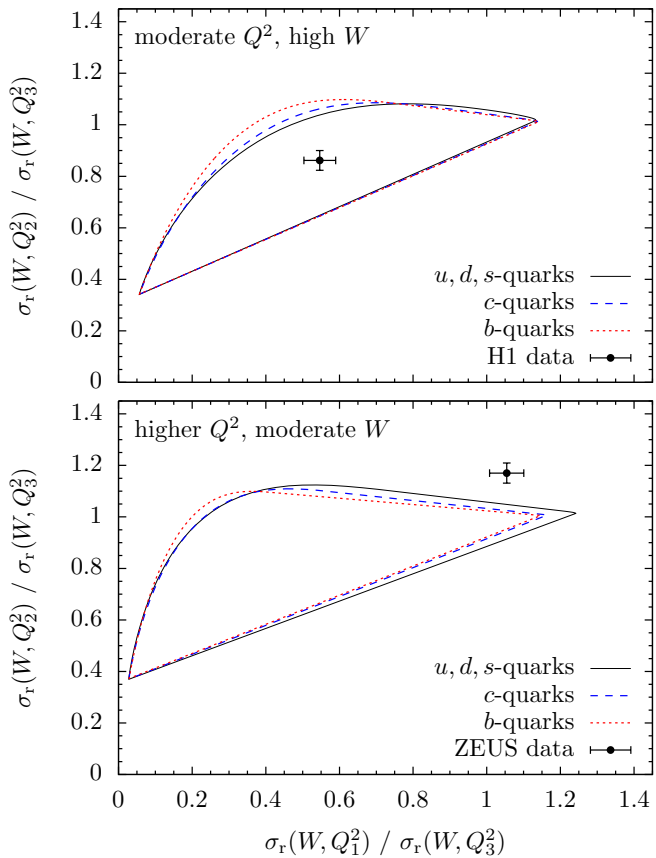}
\caption{
\label{fig-redbndshera}
Correlated bounds on ratios of the reduced cross section 
\eqref{sigreddef} at different values of $Q^2$ obtained from
the dipole picture, confronted with HERA data. 
The inner parts of the different curves show the allowed regions if only 
contributions from specific quark flavors are considered, while the 
convex hull of these regions gives the net bound if no further assumptions 
are made. The kinematical values are 
$W=247$~GeV, $(Q_1^2,Q_2^2,Q_3^2)=(2,12,35)\text{~GeV}^2$ 
for the upper plot and $W=75$~GeV, 
$(Q_1^2,Q_2^2,Q_3^2)=(3.5,45,120)\text{~GeV}^2$ for the lower plot. 
}
\end{figure}

In the following we want to confront the bounds that we obtain 
in this section from the color dipole picture with HERA data. 
Before we proceed with that a remark is in order concerning 
that comparison with data. 
Data on $F_2$ and measurements of the reduced cross 
section are available for a large range of $Q^2$ values with 
$(x,Q^2)$-binning. However, throughout this section we deal 
with bounds involving values of $F_2$ (or of the reduced cross section, 
see below) at the same $W$ but at different $Q_i^2$. Hence a comparison 
with our bounds requires different values of $Q^2$ at the same value of $W$, 
and data with $(W,Q^2)$-binning 
are published only for comparatively small kinematical ranges. 
We therefore use a fit to the $F_2$ data that can, to a good approximation, 
be considered as a substitute of actual data. We do this in most 
of the following comparisons, except for two illustrations where HERA data 
are used directly (see Fig.\ \ref{fig-redbndshera} and the corresponding 
discussion below). Concretely, we use the ALLM97 fit to $F_2$ 
\cite{Abramowicz:1991xz,Abramowicz:1997ms} which represents the 
measured data points of \cite{Breitweg:2000yn}-\cite{Chekanov:2003yv} 
within their errors, except maybe for the 
region of very low $Q^2$ where the fit appears to be slightly worse. 
We emphasize that we use the fit only {\em inside} the kinematical 
range in which actual HERA data are available. No extrapolation 
beyond that range is done here. 

\begin{figure}
\includegraphics[width=\columnwidth]{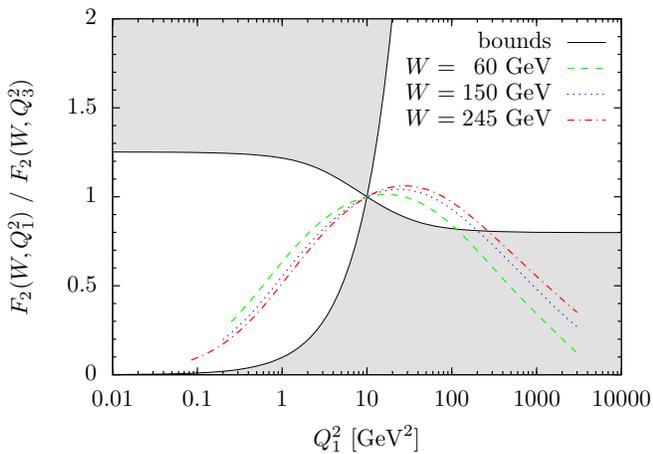}
\caption{
\label{fig-f2allm97}
The bounds \eqref{f2uncbnds13} on $F_2(W,Q_1^2)/F_2(W,Q_3^2)$ 
resulting from the dipole picture (solid lines) confronted with the 
corresponding ratios obtained using the ALLM97 fit to $F_2$ for 
three different values of $W$. 
Here $Q_1^2$ is varied while the value $Q_3^2=10\text{~GeV}^2$ 
is kept fixed. The shaded region is excluded by the bounds. 
}
\end{figure}

Fig.\ \ref{fig-f2allm97} confronts the bound \eqref{f2uncbnds13} with 
the ALLM97 fit to $F_2$ for a fixed value of $Q_3^2$ and variation 
of $Q_1^2$, as presented in \cite{Ewerz:2006an} before. 
It is apparent from the Figure that there is a value of $Q_1^2$ beyond 
which the dipole picture fails to be compatible with the ALLM97 fit. 
This maximal $Q_1^2$ value depends on the value of $W$, as can be 
seen in the Figure from the three curves for different $W$, 
and it also depends on the value chosen for $Q_3^2$. With the 
choice $Q_3^2=10\,\mbox{GeV}^2$ made for the Figure 
this maximal $Q_1^2$ is in the range 
of about $150$-$300\,\mbox{GeV}^2$, depending on $W$. 

\begin{figure}
\includegraphics[width=\columnwidth]{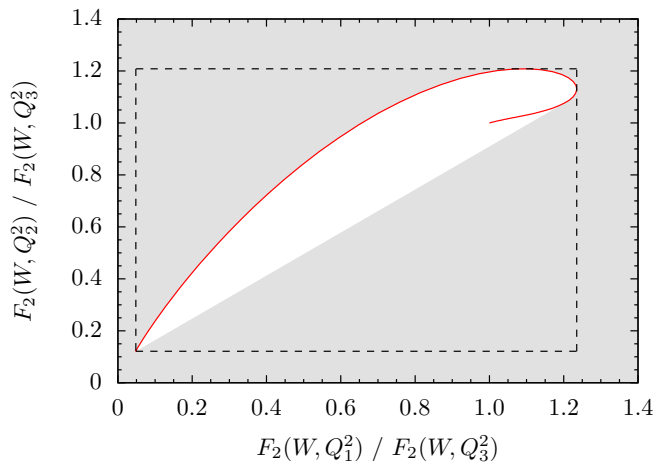}
\caption{
\label{figtrajectoryproj}
The trajectory of the vectors 
$(f^{(q)}(r,Q_1^2)/f^{(q)}(r,Q_3^2), f^{(q)}(r,Q_2^2)/f^{(q)}(r,Q_3^2))^{\rm T}$ 
for variation of $r$ (solid curve), here for a massless quark flavor $q$ 
and for the choice $(Q_1^2,Q_2^2,Q_3^2)=(4,10,80)\text{~GeV}^2$. 
The unshaded area is the closed convex hull of the vectors that form 
the trajectory. According to the weaker bounds 
\eqref{f2uncbnds13} and \eqref{f2uncbnds23} the vectors 
$(F_2(W,Q_1^2)/F_2(W,Q_3^2), F_2(W,Q_2^2)/F_2(W,Q_3^2))^{\rm T}$ 
must lie within the dashed rectangle, while the stronger bound 
\eqref{f2corbnds} requires them to lie within the convex hull 
of the unshaded area and the corresponding areas obtained for 
massive quarks. The curves for massive quarks have a similar shape 
and are not shown here for simplicity. 
}
\end{figure}

So far we have discussed in some detail the bounds \eqref{f2uncbnds13} 
and \eqref{f2uncbnds23} which resulted from considering the two 
components of \eqref{f2corbnds} separately. We can improve these 
bounds by taking into account the correlation of those two components, 
that is the correlation of the two ratios 
$F_2(W,Q_1^2)/F_2(W,Q_3^2)$ and $F_2(W,Q_2^2)/F_2(W,Q_3^2)$. 
According to \eqref{f2corbnds} the 2-vector constructed of these two 
ratios for a given set of $Q_i^2$ lies in the closed convex hull
of the vectors $(f^{(q)}(r,Q_1^2)/f^{(q)}(r,Q_3^2), 
f^{(q)}(r,Q_2^2)/f^{(q)}(r,Q_3^2))^{\rm T}$. 
As $r$ is varied the latter vector (for each quark flavor $q$) 
follows a trajectory in 2-dimensional space. 
For the case of a massless quark flavor $q$ that trajectory is shown 
as the solid curve in Fig.~\ref{figtrajectoryproj}, where we have chosen 
the values $(Q_1^2,Q_2^2,Q_3^2)=(4,10,80)\text{~GeV}^2$ for this 
example. The white (unshaded) area is the closed convex hull of the vectors 
that form the trajectory. Similar but slightly different trajectories 
are obtained for massive quark flavors, which are not shown here 
in order to keep the figure simple. 
As a consequence of the dipole picture the 2-vectors 
$(F_2(W,Q_1^2)/F_2(W,Q_3^2), F_2(W,Q_2^2)/F_2(W,Q_3^2))^{\rm T}$ 
must lie within the closed convex hull of those trajectories,
independently of the energy $W$, see \eqref{f2corbnds}. 
The dashed lines in Fig.\ \ref{figtrajectoryproj} represent the 
two bounds \eqref{f2uncbnds13} and \eqref{f2uncbnds23}. 
Clearly, the correlated bound \eqref{f2corbnds} is much 
stronger than those separate bounds on the ratios. 

Next we want to compare the stronger bound \eqref{f2corbnds} 
with experimental data. For this purpose we need data points of $F_2$ at 
three different $Q_i^2$ but at the same $W$. However, most of the 
available data are not published in $(W,Q^2)$-binning. We have 
found only few points which are suitable for a direct comparison 
with our bound, that is with the same $W$ and three different $Q_i^2$ 
that are not too close to each other. 
We will now present two of these examples. 
Further below we will then again use the ALLM97 fit for a 
more comprehensive analysis of the kinematical range in which the 
bound  \eqref{f2corbnds} is respected. 
For the comparison with actual HERA data we choose as 
the observable the reduced cross section instead of $F_2$, 
since the former is the one which was 
directly measured. The reduced cross section is defined as 
\begin{equation}
\label{sigreddef}
\sigmared (W,Q^2) =
 \frac{Q^2}{4 \pi^2 \alpha_{\text{em}}}
  \left( \sigmatot_T
  + \frac{2(1-y)}{1+(1-y)^2} \sigmatot_L \right),
\end{equation}
with $y \approx (W^2 + Q^2)/s$, see \eqref{defkin}, where
$\sqrt{s}\approx 300\text{~GeV}$ is the lepton-proton
center-of-mass energy for the available HERA data. 
It is straightforward to derive correlated bounds for ratios of reduced 
cross sections instead of $F_2$ structure functions from the dipole picture. 
The derivation is completely analogous to the one described above. 
We just have to replace $f^{(q)}$ by 
\begin{equation}
\label{defftilde}
f_{\rm r}^{(q)} = 
Q^2 \left[w_T^{(q)}(r,Q^2)
+\frac{2(1-y)}{1+(1-y)^2} w_L^{(q)}(r,Q^2)\right]\,,
\end{equation}
as can be seen from \eqref{f2dipsimple} and \eqref{sigreddef} 
together with \eqref{flc2vec}, \eqref{f2fdef}. The resulting bound 
is then as given by \eqref{f2corbnds} but with $F_2$ replaced by 
$\sigmared$ and $f^{(q)}$ replaced by $f_{\rm r}^{(q)}$. 
Due to this modification the bound for the reduced cross section now 
depends on $W$ (which enters via $y$), 
which was not the case for the original bound for $F_2$. 
Fig.~\ref{fig-redbndshera} confronts the bound on the quantity 
$(\sigmared(W,Q_1^2)/\sigmared(W,Q_3^2), 
\sigmared(W,Q_2^2)/\sigmared(W,Q_3^2))^{\rm T}$ 
with its measured values from ZEUS \cite{Chekanov:2001qu} and 
H1 \cite{Adloff:2000qk} for two different choices of $W$ and of 
the triple of $Q_i^2$. 
The depicted errors on the ratios are the combination of the 
experimental errors on $\sigmared$ in quadrature. 
The curves in Fig.~\ref{fig-redbndshera} show the correlated bounds 
for contributions to $(\sigmared(W,Q_1^2)/\sigmared(W,Q_3^2), 
\sigmared(W,Q_2^2)/\sigmared(W,Q_3^2))^{\rm T}$ from different quark 
flavors as given by the analog of \eqref{f2corbnds} for $\sigmared$. 
Only if the point obtained from the data lies within the convex hull 
of all these curves it can possibly be described in the framework 
of the dipole picture. 
We see that this condition is fulfilled for the high $W$, moderate 
$Q^2$ sample (upper graph), while it is violated by approximately 
two standard deviations for the lower $W$, higher $Q^2$ sample 
(lower graph).

The above discussion refers to the applicability of the dipole 
picture at a given value of $W$ for one particular triple of $Q^2$-values. 
For a determination of the range of applicability of the 
dipole picture it is more desirable to determine for a given $W$ 
a maximal {\em range} in $Q^2$ in which the three $Q_i^2$ can be 
chosen without giving rise to a violation of the bound. 
For this purpose we now consider again the structure function 
$F_2$ (and no longer the reduced cross section). 
Using the ALLM97 fit to the measured $F_2$ data 
we can then perform a continuous scan in $Q^2$ and determine 
precisely the kinematical range in which the bounds are respected. 
We first consider the correlated bounds 
obtained from \eqref{f2corbnds}, and later compare the allowed 
$Q^2$-range with the one resulting from the weaker bounds 
\eqref{f2uncbnds13} and \eqref{f2uncbnds23}. 

Let us first fix the energy $W$ at some value. We will in the following 
call a violation of the dipole-picture bound a `significant' violation if 
the ALLM97 $F_2$ ratios give a relative deviation of more than 10\% 
from the bound. This accounts for a kind of error band which should be 
associated with the ALLM97 fit or with the corresponding ratios of $F_2$. 
If for a certain triple $(Q_1^2,Q_2^2,Q_3^2)$ the ratios obtained from 
the ALLM97 fit violate the bounds by a significant amount (in the 
above sense) any $Q^2$-range containing the values 
$Q_1^2,Q_2^2,Q_3^2$ is excluded for a 
successful description within the dipole picture. 
In contrast, agreement with the bounds for a triple does not 
necessarily imply agreement for the full range 
$[\min_i(Q_i^2),\max_i(Q_i^2)]$ of that triple 
since the bounds depend on all three $Q_i^2$. 
We therefore systematically search for the maximal 
$Q^2$-range that contains no $Q^2$-triple for which 
the bounds are violated significantly. Technically, we do this 
by searching for the minimal $Q^2$-range in which 
we can find at least one $Q^2$-triple for which the bounds are 
violated significantly. 
The lower bound of a given $Q^2$-interval turns out to have only 
mild influence on whether a significant violation of the bounds 
can be found within that interval -- provided it is not 
much larger than $1\text{~GeV}^2$. 
We therefore keep the lower end of the considered $Q^2$-range 
fixed at $1\text{~GeV}^2$ and determine the upper end $Q_{\rm max}^2$ 
of the $Q^2$-range within which the bounds are not significantly 
violated. We can then repeat this procedure for each energy $W$ 
and determine $Q_{\rm max}^2$ as a function of $W$. 

\begin{figure}
\includegraphics[width=\columnwidth]{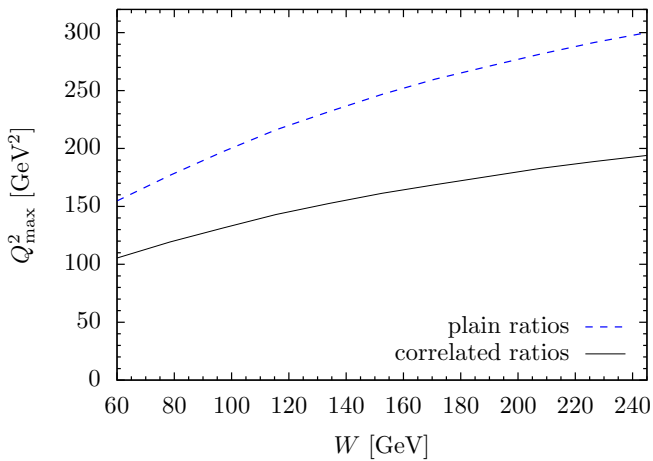}
\caption{
\label{fig-maxq2}
Upper limit $Q_{\rm max}^2$ of the $Q^2$-range in which 
the ALLM97 fit to $F_2$ is consistent with the bounds obtained 
from the dipole picture 
within a 10\% relative deviation of the $F_2$ ratios, plotted 
as a function of the energy $W$. 
The curves are for consistency with the correlated bounds 
\eqref{f2corbnds} (solid line), and with the uncorrelated bounds 
\eqref{f2uncbnds13} and \eqref{f2uncbnds23} (dashed line), 
respectively. 
}
\end{figure}

The solid line in Fig.~\ref{fig-maxq2} shows the result of such 
a calculation based on the correlated bounds obtained from 
\eqref{f2corbnds}. The allowed $Q^2$-range slowly grows with 
increasing energy, as can be expected on general grounds. 
$Q_{\rm max}^2$ ranges from about $100\,\mbox{GeV}^2$ 
for $W=60\,\mbox{GeV}$ to about $200\,\mbox{GeV}^2$ 
for $W=245\,\mbox{GeV}$. 
The dashed line in Fig.~\ref{fig-maxq2} represents the 
analogous curve obtained from the uncorrelated bounds 
\eqref{f2uncbnds13} and \eqref{f2uncbnds23}. Here 
we have varied both $Q_1^2$ and $Q_3^2$ in \eqref{f2uncbnds13} 
in order to determine the maximal virtuality, $Q_{\rm max}^2$, 
below which both $Q_1^2$ and $Q_3^2$ can be chosen arbitrarily 
without giving rise to a significant violation of the bound. 
We see that the correlated bounds resulting from 
\eqref{f2corbnds} indeed give stronger restrictions 
on the range of validity of the dipole picture than 
the uncorrelated bounds \eqref{f2uncbnds13}, \eqref{f2uncbnds23}. 

Note that the violation of the correlated bound does 
not take place at a constant value of $x$. In the contrary, 
the value of $x$ changes along the solid line in Fig.\ \ref{fig-maxq2}. 
For $Q^2=100 \,\text{GeV}^2$ we find that $x<0.03$ is required for 
the bound not to be violated, while for $Q^2=200 \,\text{GeV}^2$ 
the bound is only respected for $x<0.003$. 
A similar observation applies to the uncorrelated bounds 
\eqref{f2uncbnds13}, \eqref{f2uncbnds23} (see the dashed line 
in Fig.\ \ref{fig-maxq2}) as already observed in \cite{Ewerz:2006an}. 

Obviously, a violation of the above bounds indicates that some contributions 
to the cross section become important which are not contained 
in the dipole picture. We would like to emphasize that such 
corrections to the dipole picture might become sizable already before 
the bounds are actually violated. One should therefore expect that 
corrections to the standard dipole picture are important already if 
the data come close to the bounds.  

The upper limit on the kinematical range of validity of the 
dipole picture that we find here appears to be rather low 
in view of the fact that phenomenological fits to $F_2$ data based 
on the dipole picture often work quite well up to rather high $Q^2$, 
see for example \cite{Bartels:2002cj}. However, the good quality 
of those fits at large $Q^2$ is not in contradiction with our result. 
We recall that the bounds derived in this section crucially 
depend on the correct functional dependence of the dipole 
cross section $\sigmadip$ on $r$ and $W$, as obtained 
naturally from the derivation of the dipole picture 
presented in \cite{Ewerz:2004vf,Ewerz:2006vd}. 
In particular, $\sigmadip$ needs to be independent of $Q^2$ 
for our bounds to be valid. But almost all recent models 
for $\sigmadip$ assume it to depend on $x$, and hence on 
$Q^2$. The transition from the energy variable $W$ to 
the energy variable $x$ in the dipole cross section 
requires additional assumptions the justification and the 
physical significance of which appears difficult to assess. 
In practice, they might 
capture -- at least partly -- some corrections that are left 
out in the usual dipole picture (see the discussion in section 
\ref{sec-dipole}). It would be very desirable to obtain a 
better understanding of this situation. An important 
step would be to check whether it is also possible to 
describe the presently available HERA data by 
models for the dipole cross section based on the 
more natural variables $r$ and $W$.

We finally note that our bounds are modified if one uses, 
as suggested by Hand's convention, the relation 
\eqref{f2diphand} between $F_2$ and the $\gamma^* p$ 
cross sections $\sigma_T$ and $\sigma_L$ instead of 
the simpler relation \eqref{f2dipsimple} that has been 
used here. The modification becomes relevant for 
not so small $x$, and hence for large $Q^2$. The 
natural kinematical region for the application of 
the dipole picture is the region of small $x$, so that 
this modification is only of minor relevance for 
the dipole picture. Nevertheless, we find it interesting to see 
the effect of using \eqref{f2diphand} on our bounds and illustrate it in 
an example in appendix~\ref{app-hand}. Our conclusions 
about the range of validity of the dipole picture would 
not to be significantly affected. 

\section{Conclusions}
\label{sec-concl}

The dipole picture of high energy photon-proton scattering is a 
popular framework for the analysis and interpretation of HERA data. 
However, the dipole picture is not exact, and a number of assumptions 
and approximations are needed to obtain it from the general description 
of photon-proton scattering. It is therefore important to determine 
as precisely as possible its kinematical range of validity. Using the dipole 
picture beyond that range could clearly result in misleading conclusions. 

In the present paper we have briefly summarized the assumptions 
and approximations underlying the dipole picture in addition to 
taking the high energy limit. In particular we have indicated 
some contributions to the $\gamma^* p$ cross section which 
are not contained in the dipole picture and might give rise 
to significant correction terms in some kinematical regions. 
We have then derived various bounds 
on ratios of deep inelastic structure functions from the dipole picture. 
These bounds involve only the photon wave functions and do not make 
use of any model assumptions about the dipole-proton cross section. 
They have to be respected in the kinematical range of applicability 
of the dipole picture. A comparison with experimental data then 
allowed us to constrain this range independently of any model assumptions. 

We have first considered the structure functions $F_L$, $F_2^{(c)}$, 
and $F_2$, all taken at the same $W$ and $Q^2$. 
From the dipole picture we have obtained an upper bound on 
$F_L/F_2$ as well as a correlated upper bound on the ratios 
$F_L/F_2$ and $F_2^{(c)}/F_2$. It will be interesting to compare 
these bounds with future results from measurements of these 
structure functions. 

Furthermore, we have derived correlated bounds on ratios of 
$F_2$ at three different $Q_i^2$ but at the same energy $W$. 
These bounds are significantly more restrictive than bounds 
on simple ratios that had been obtained already in \cite{Ewerz:2006an}. 
We have compared these bounds with experimental 
data. More precisely, we have used the ALLM97 fit to the 
measured $F_2$ data except for two examples in which we have used 
actual data points. Since our bounds apply to ratios of $F_2$ at the 
same $W$ a more direct comparison would require to have the data in 
$(W,Q^2)$-binning instead of the commonly used $(x,Q^2)$ binning. 
Employing the ALLM97 fit within the kinematical range of HERA 
we have computed ranges in $Q^2$ in which the bounds obtained 
from the dipole picture are respected. We have further studied the 
dependence of these ranges on the energy $W$. Depending on $W$ the 
dipole picture fails to be applicable above a $Q^2$ of about $100 $ to 
$200\,\mbox{GeV}^2$. We expect that already for values of 
$Q^2$ somewhat below those limits corrections to the 
usual dipole picture become important. For low $Q^2$ the bounds 
on ratios of $F_2$ are found to be respected by the data. 

We should point out that the bounds on ratios of $F_L$, $F_2^{(c)}$ 
and $F_2$ obtained in section \ref{sec-flc2} are independent of the 
choice of energy variable in the dipole cross section $\sigmadip$. 
The bounds on ratios of $F_2$ at the same $W$ but different 
$Q_i^2$ discussed in section \ref{sec-f2q2}, on the other hand, 
crucially require that $\sigmadip$ is independent of $Q^2$. 
Similarly, other modifications of the standard dipole picture formulae 
\eqref{denst}-\eqref{sigmaldip} might in general affect the bounds 
resulting from these formulae. For an example see 
\cite{Schildknecht:2007wg}. 

We recall that in \cite{Ewerz:2006vd,Ewerz:2006an} an 
upper bound on the ratio $R=\sigma_L/\sigma_T$ has been 
derived from the dipole picture, $R \le 0.372$. 
The experimental data for that ratio have large errors but appear 
to come close to the bound for $Q^2$ below about $2\,\mbox{GeV}^2$, 
which can be interpreted as a possible breakdown of the dipole picture 
in this region of low $Q^2$. Combining this with the findings of the 
present paper we conclude that the data are compatible with the 
bounds resulting from 
the dipole picture for $Q^2$ between $2\,\mbox{GeV}^2$ and 
$100$-$200\,\mbox{GeV}^2$, with the upper limit depending on $W$. 
Any results that depend on using the dipole picture outside this 
kinematical range might be considerably affected by potential 
corrections to the standard dipole picture and should be interpreted 
only with great care. 

\begin{acknowledgments}
We thank M.\ Diehl, J.\ Forshaw, H.\ Kowalski, M.\ Ryskin, 
D.\ Schildknecht, and G.\ Shaw for useful discussions. 
This work was supported by the Deutsche Forschungsgemeinschaft, 
project number NA 296/4-1. 
\end{acknowledgments}

\appendix

\boldmath
\section{The Normalization of $\gamma^* p$ Cross Sections and its Consequences in the Dipole Picture}
\unboldmath
\label{app-hand}

In this appendix we would like to illustrate in one example how our 
bounds are affected by different normalizations of the 
$\gamma^* p$ cross sections $\sigma_T$ and $\sigma_L$ relative 
to $F_2$. In the context of high energy scattering, in particular 
when using the dipole picture, one usually employs the simple 
relation \eqref{f2dipsimple}. The relation \eqref{f2diphand} 
derived from Hand's convention reduces to that simple 
expression if the high energy limit is taken for fixed $Q^2$. 
For finite $x$ the normalization of $\sigma_{T,L}$ relative 
to $F_2$ differs by a factor $(1- x)$ between the two choices 
(neglecting terms of ${\cal O}(m_p^2/W^2)$), as is 
indicated in \eqref{f2diphand}. 
\begin{figure}[tb]
\includegraphics[width=\columnwidth]{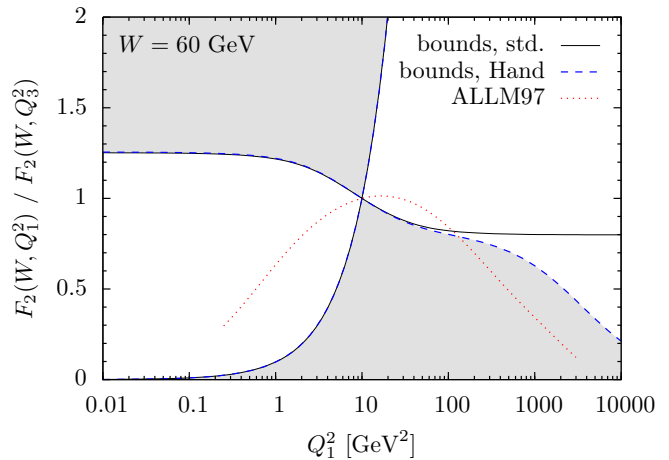}
\caption{
\label{figchangeHand}
Change of the bound \eqref{f2uncbnds13} 
on $F_2(W, Q_1^2)/F_2(W, Q_3^2)$ due to using 
Hand's convention \eqref{f2diphand} instead of the 
simpler \eqref{f2dipsimple}, shown here 
for $W=60\,\mbox{GeV}$ and the choice $Q_3^2=10\,\mbox{GeV}^2$. 
The solid lines are the original bounds \eqref{f2uncbnds13}, while the dashed 
lines represent the modified bounds. 
The dotted line is the ALLM97 fit to 
$F_2(W, Q_1^2)/F_2(W, Q_3^2)$. 
}
\end{figure}

The bounds discussed in the present paper are all based on 
the simpler formula \eqref{f2dipsimple}. It is straightforward, 
however, to derive similar bounds based on the relation 
\eqref{f2diphand}. The additional factor $(1-x)$ depends 
both on $Q^2$ and $W$. Therefore this factor does not 
cancel if ratios of structure functions are taken at different 
$Q_i^2$. Furthermore, the bounds on ratios of $F_2$ 
inherit a dependence on $W$ from this factor. 
The bounds on ratios of $F_L$, $F_2^{(c)}$, and $F_2$ 
discussed in section \ref{sec-flc2}, on the other hand, 
are not affected. There the structure functions are 
evaluated at the same $W$ and $Q^2$ and the additional factor 
$(1-x)$ cancels in the ratios. 

In Fig.\ \ref{figchangeHand} we show for one energy 
$W=60\,\mbox{GeV}$ how the bound \eqref{f2uncbnds13} is
 changed if one uses \eqref{f2diphand} instead of \eqref{f2dipsimple}. 
A sizable deviation from the original bound occurs only at relatively 
large $Q^2$ where the new bound is closer to the data than the 
original bound. However, both bounds are violated by the data 
at about the same $Q^2$ and the difference between the 
original and the modified bound grows only at larger $Q^2$. 
Similar remarks apply to the correlated bounds on ratios 
of $F_2$. Therefore the normalization of the $\gamma^* p$ 
cross sections according to Hand's convention would not 
significantly alter our results concerning the range of 
validity of the dipole picture. 

\section{Convex~Hulls, Convex~Cones and Moment~Problems}
\label{app-convex}

In this appendix we discuss the notions of convex hull and convex cone as well
as some further mathematical relations.
The precise mathematical definitions can be found in \cite{encyclo},
for our notation see also \cite{schoel}.

Let us consider the $n$-dimensional Euclidean space $\realnums^n$ with
elements $\nvec x$, $\nvec y$ etc.
A non-empty subset $X$ of $\realnums^n$ is called a {\em convex set} if for
any elements $\nvec x$, $\nvec y$ in $X$ and any real number $a$ with
$0\leq a\leq 1$ the element $a\nvec x+(1-a)\nvec y$ is also contained in $X$.
That is, with any two points of $X$ the complete straight line connecting them
is also in $X$.

Let now $Y$ be an arbitrary nonempty subset of $\realnums^n$.
The minimum convex set containing $Y$ exists~\cite{encyclo} and is called
the {\em convex hull} of $Y$ and denoted by $\cohull(Y)$.
Its closure is denoted by $\cohullcl(Y)$.
To illustrate this concept we give a physical example.
Let $Y=\{\nvec y^{(1)},\dots, \nvec y^{(N)}\}$ be a set of $N$ points in
$\realnums^n$.
Consider arbitrary distributions of masses $m_i\geq 0$ $(i=1,\dots, N)$ on
these points.
The center of mass is then
\begin{equation}\label{B.1}
\nvec{x} = \frac{ \sum_{i=1}^{N} m_i \nvec{y}^{(i)} }{ \sum_{i=1}^{N} m_i }\,.
\end{equation}
The convex hull of $Y$, $\cohull(Y)$, is the set of all possible center of
mass points of such mass distributions.

Next we discuss the notion of {\em convex cone}.
A nonempty subset $X$ of $\realnums^n$ is called a convex cone if for any
elements $\nvec x$, $\nvec y$ of $X$ and any real number $a\geq 0$
the elements $a\nvec x$ and $\nvec x+\nvec y$ are also contained in $X$.
Let $Y$ be an arbitrary non-empty subset of $\realnums^n$, then the minimal
convex cone containing $Y$ exists and is denoted by $\cocone(Y)$.
Its closure is denoted by $\coconecl(Y)$.

\begin{figure}[tb]
\centering
\includegraphics[width=0.65\columnwidth]{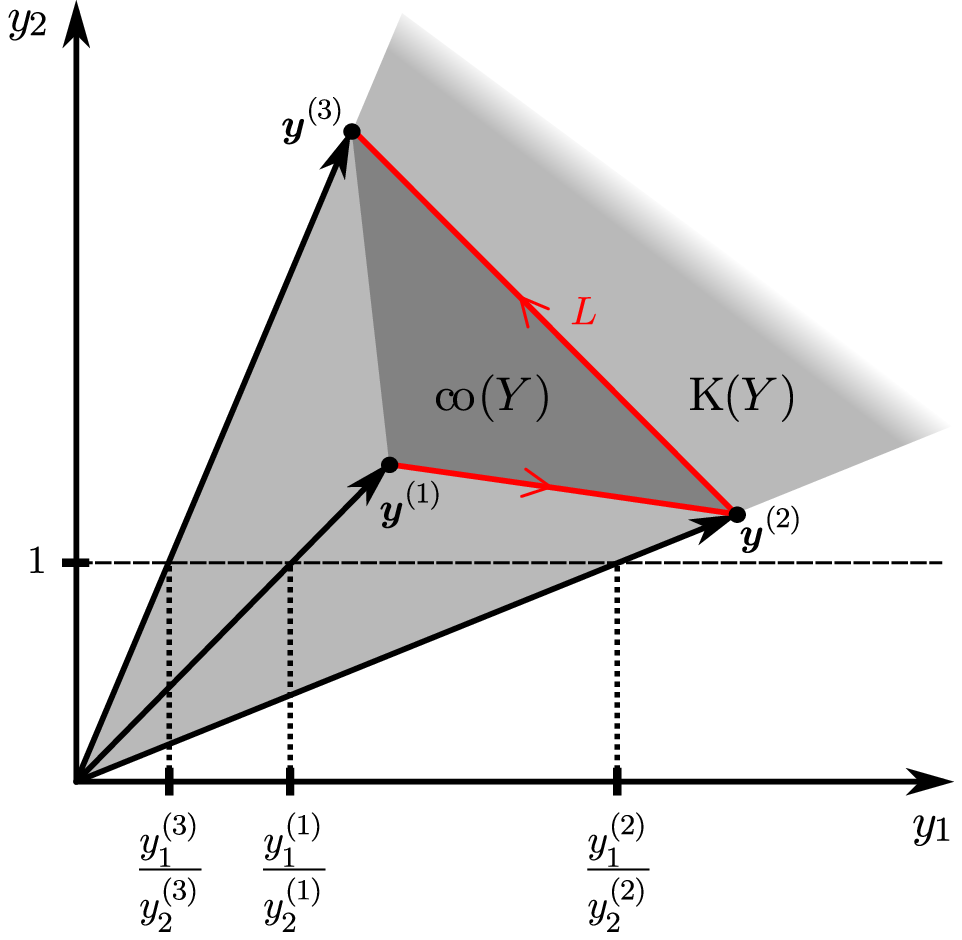}
\caption{\label{figconvex}
Illustration of the convex hull $\cohull(Y)$ and the convex cone $\cocone(Y)$
for the set $Y$ of \eqref{B.2}.
Here both sets are closed, that is $\cohullcl(Y)=\cohull(Y)$ and
$\coconecl(Y)=\cocone(Y)$.}
\end{figure}

We illustrate these notions with a two-dimensional example.
Let $Y$ consist of three points in $\realnums^2$
\begin{equation}\label{B.2}
Y = \left\{ \nvec{y}^{(1)}, \nvec{y}^{(2)}, \nvec{y}^{(3)} \right\}
\end{equation}
as shown in Fig.~\ref{figconvex}.
Here
\begin{equation}\label{B.3}
\nvec{y}^{(i)} = \begin{pmatrix} y^{(i)}_1 \\ y^{(i)}_2 \end{pmatrix}
\end{equation}
and we suppose 
\begin{equation}\label{B.4}
y^{(i)}_2 > 0\qquad\text{for~} i=1,2,3\,.
\end{equation}
The convex hull of $Y$, $\cohull(Y)$, is given by the dark grey triangle
bounded by the polygon from $\nvec y^{(1)}$ to $\nvec y^{(2)}$, $\nvec y^{(3)}$
and back to $\nvec y^{(1)}$.
The cone $\cocone(Y)$ is indicated by the light grey area bounded by the rays
$\lambda\,\nvec y^{(2)}$ with $\lambda\geq 0$ and $\mu\,\nvec y^{(3)}$ with
$\mu\geq 0$. 

Let $Y$ be a non-empty subset of $\realnums^n$ and let $\cohull(Y)$ be the
convex hull of $Y$ and $\cocone(Y)$ the minimal convex cone containing $Y$.
We define the set
\begin{equation}\label{B.4a}
\cocone'(Y) = \left\{ \left. \nvec{x} \,\right\vert~ \nvec{x} = \lambda \nvec{y},~
\lambda \ge 0,~\nvec{y}\in\cohull(Y)\,\right\}
\end{equation}
and assert that
\begin{equation}\label{B.4b}
\cocone'(Y) = \cocone(Y)\,.
\end{equation}
The proof of \eqref{B.4b} goes as follows.
It is easy to see that $\cocone'(Y)$ is a convex cone containing $Y$.
Thus, since $\cocone(Y)$ is the minimal such cone we have
\begin{equation}
\cocone'(Y) \supset \cocone(Y)\,.
\end{equation}
On the other hand, $\cocone(Y)$ is a convex set containing $Y$ and $\cohull(Y)$
is the minimal such set. Thus
\begin{equation}\label{B.4d}
\cocone(Y) \supset \cohull(Y)\,.
\end{equation}
Since $\cocone(Y)$ is a convex cone this implies that for any element
$\nvec{x}\in\cohull(Y)$ and any $\lambda\ge 0$ we have
$\lambda\nvec{x}\in\cocone(Y)$.
That is, we have
\begin{equation}\label{B.4e}
\cocone'(Y) \subset \cocone(Y) \,.
\end{equation}
Therefore, we have shown that $\cocone'(Y)=\cocone(Y)$.
That is, every element of $\cocone(Y)$ can be written in the form
given in \eqref{B.4a}.
For the closures we find in a similar way
\begin{equation}\label{B.4f}
\coconecl(Y) = \left\{ \left. \nvec{x} \,\right\vert~
 \nvec{x} = \lambda\nvec{y},~ \lambda\ge 0,~\nvec{y}\in\cohullcl(Y)\,\right\}\,.
\end{equation}

Next we come to the moment problem which is at the heart of our derivations
of bounds.
Suppose we have a continuous vector function on a closed interval
$[t_0,t_1]\subset\realnums$ defining a curve $L$ in $\realnums^n$:
\begin{align}\label{B.5}
L:\quad [t_0,t_1] &\rightarrow \realnums^n\,,\notag\\
t &\mapsto \nvec{y}(t)\,.
\end{align}
We also suppose that there is at least one constant vector $\nvec{a}$ such that
\begin{equation}\label{B.5a}
\nvec{a}^{\mathrm{T}} \nvec{y}(t) > 0\qquad \text{for all~} t \in [t_0,t_1]\,.
\end{equation}
We are interested in the set $\tilde K$ of all points $\nvec x$ of
$\realnums^n$ which can be represented as
\begin{equation}\label{B.6}
\nvec{x} = \int_{t_0}^{t_1}\! \nvec{y}(t)\, \ud\Sigma(t)
\end{equation}
where $\Sigma(t)$ is some non-decreasing function on $[t_0,t_1]$.
Note that such a function is bounded from below and above since
\begin{equation}\label{B.6a}
\Sigma(t_0) \leq \Sigma(t) \leq \Sigma(t_1)\,.
\end{equation}
Before we discuss the solution of this problem as given in \cite{ahierzer} we
note that in \eqref{B.6} we are dealing with so-called Stieltjes integrals,
see for example \cite{smirnov,mathdict}.
The reader not familiar with these integrals may always set
\begin{equation}\label{B.7}
\ud\Sigma(t) = \sigma(t)\,\ud{t}
\end{equation}
where $\sigma(t)$ is some non-negative distribution.
That is, $\sigma(t)$ can be an ordinary non-negative function but can also
contain non-negative $\delta$-distributions.

The solution of the problem posed above is as follows, see \cite{ahierzer}.
The set $\tilde K$ of points which can be represented in the form \eqref{B.6}
is given by $\coconecl(L)$, that is, by the smallest closed convex cone
containing the curve $L$:
\begin{equation}\label{B.8}
\tilde{K} = \coconecl(L)\,.
\end{equation}

Consider for illustration the two-dimensional example as in
Fig.~\ref{figconvex} and the following curve $L$ defined for $t\in [0,1]$,
\begin{equation}\label{B.9}
L:~ t \mapsto \nvec{y}(t) =\!\!\begin{cases}
  \nvec{y}^{(1)} + 2 t (\nvec{y}^{(2)}-\nvec{y}^{(1)})&
    \!\!\!: 0\le t\le \frac{1}{2}\,,\\
  \nvec{y}^{(2)} +(2t-1)(\nvec{y}^{(3)}-\nvec{y}^{(2)})&
    \!\!\!: \frac{1}{2}< t\le 1\,.
  \end{cases}
\end{equation}
We are interested in the points $\nvec x$ allowing a representation
\begin{equation}\label{B.10}
\nvec{x} = \int_0^1\! \nvec{y}(t)\, \ud\Sigma(t)
\end{equation}
with some non-decreasing function $\Sigma(t)$.
According to the theorem quoted above $\nvec x$ has to be in the closed
convex cone $\coconecl(L)$ as shown in Fig.~\ref{figconvex}.
We ask now for the allowed range for $x_1$ given some $x_2$.
The possible $x_1$ values are obtained by cutting the cone $\coconecl(L)$ at
$y_2=x_2=\text{\it const.}$ and reading off the corresponding $y_1$ values.
Similarly, the allowed range of the ratio $x_1/x_2$ is obtained if we
choose $y_2=1$ for cutting the cone.
Clearly, to get the extremal values of $x_1/x_2$ we just have to consider the
generating rays $\lambda\nvec y^{(i)}$, $\lambda\geq 0$, for $i=1,2,3$.
Cutting them at $y_2=1$ gives $y^{(i)}_1/y^{(i)}_2$, $i=1,2,3$.
Among these ratios there are the extremal points of $x_1/x_2$.
In our example we get
\begin{equation}\label{B.11}
\frac{y_1^{(3)}}{y_2^{(3)}} \le \frac{x_1}{x_2} \le
\frac{y_1^{(2)}}{y_2^{(2)}}\,.
\end{equation}
Note that the interval $[y^{(3)}_1/y^{(3)}_2~,~y^{(2)}_1/y^{(2)}_2]$ is the
convex hull of the set $\{y^{(i)}_1/y^{(i)}_2 \,\vert\, i=1,2,3\}$.

For the general case, in $\realnums^n$, the situation is completely analogous.
Consider \eqref{B.6} in $\realnums^n$ $(n\geq 2)$ and let us write in
components
\begin{equation}\label{B.12}
\begin{pmatrix} x_1 \\ \vdots \\ x_n \end{pmatrix} =
  \int_{t_0}^{t_1} \begin{pmatrix} y_1(t) \\ \vdots \\ y_n(t) \end{pmatrix}
  \ud\Sigma(t)\,.
\end{equation}
Let us suppose that 
\begin{equation}\label{B.12a}
0 < c_0 \le y_n(t) \le c_1 \qquad \text{for~}t_0 \le t \le t_1\,.
\end{equation}
Thereby \eqref{B.5a} is satisfied with $\nvec{a}^{\mathrm{T}}=(0,\ldots,0,1)$.
To get the bounds for the ratio vector
\begin{equation}\label{B.12b}
\nvec{x}' = \begin{pmatrix} x_1/x_n\\ \vdots\\ x_{n-1}/x_n\\ 1 \end{pmatrix}
\end{equation}
we just have to cut the cone $\coconecl(L)$ with the hyperplane $x_n=1$.
The corresponding set in $\realnums^n$ obtained by this cutting is given by
the closed convex hull of the ratio vectors of the curve $L$ generating the
cone $\coconecl(L)$.
That is, we denote by $L'$ the following curve in $\realnums^n$
\begin{equation}\label{B.13}
L':\quad t \mapsto \begin{pmatrix}
  y_1(t)/y_n(t)\\ \vdots\\ y_{n-1}(t)/y_n(t)\\ 1 \end{pmatrix}\,,
\qquad t \in [t_0,t_1].
\end{equation}
Let $\cohullcl(L')$ be the closed convex hull of $L'$.
The intersetion of $\coconecl(L)$ with the hyperplane $x_n=1$ is given by
$\cohullcl(L')$.
Clearly, the extremal points of the above cone--hyperplane intersection must
be given by the intersections of the rays generating the cone $\coconecl(L)$,
that is by the rays through the curve $L$. But this gives just $L'$.

We give now a formal proof of the above statements.
For this consider the minimal closed convex cone $\coconecl(L)$ containing $L$
and analogously $\coconecl(L')$ containing $L'$. We assert that 
\begin{equation}\label{B.14}
\coconecl(L) = \coconecl(L')\,.
\end{equation}
To prove \eqref{B.14} we note that according to \eqref{B.6} and \eqref{B.13} all vectors $\nvec x'\in \coconecl(L')$ are of the form
\begin{equation}\label{B.15}
\nvec{x}' = \int_{t_0}^{t_1} \frac{\nvec{y}(t)}{y_n(t)}\, \ud\Sigma'(t)
\end{equation}
with some non-decreasing function $\Sigma'(t)$. Due to \eqref{B.12a} the division by $y_n(t)$ in \eqref{B.15} is harmless and we can define a non-decreasing function $\Sigma(t)$ on $[t_0,t_1]$ by 
\begin{equation}\label{B.16}
\Sigma(t) = \int_{t_0}^t\! \frac{1}{y_n(t')}\, \ud\Sigma'(t')\,.
\end{equation}
We get then for $\nvec x'$ of \eqref{B.15}
\begin{equation}\label{B.17}
\nvec{x}' = \int_{t_0}^{t_1}\! \nvec{y}(t)\, \ud\Sigma(t)\,.
\end{equation}
That is, $\nvec x'\in \coconecl(L)$ according to \eqref{B.6} and we have shown
\begin{equation}\label{B.18}
\coconecl(L') \subset \coconecl(L)\,.
\end{equation}
Now we consider an arbitrary element $\nvec x\in\coconecl(L)$ which
according to \eqref{B.6} has the form
\begin{equation}\label{B.19}
\nvec{x} = \int_{t_0}^{t_1} \nvec{y}(t)\, \ud\Sigma(t)
\end{equation}
with some non-decreasing function $\Sigma(t)$.
We define a non-decreasing function $\Sigma'(t)$ on $[t_0,t_1]$ by 
\begin{equation}\label{B.20}
\Sigma'(t) = \int_{t_0}^t\! y_n(t')\, \ud\Sigma(t')\,.
\end{equation}
Again, we use here \eqref{B.12a}. We get then
\begin{equation}\label{B.21}
\nvec{x} =
  \int_{t_0}^{t_1}\! \nvec{y}(t) \frac{1}{y_n(t)}\,\ud\Sigma'(t)\,.
\end{equation}
That is, $\nvec x\in \coconecl(L')$ and therefore
\begin{equation}\label{B.22}
\coconecl(L) \subset \coconecl(L')\,.
\end{equation}
From \eqref{B.18} and \eqref{B.22} follows \eqref{B.14}, {\it q.e.d.}

From \eqref{B.21} we can now draw the following conclusion for any non-zero
element $\nvec x\in \coconecl(L)$.
Such an $\nvec x$ is of the form \eqref{B.19} with
$\Sigma(t) \neq \text{\it const}$.
We have then with $\Sigma'(t)$ from \eqref{B.20}, 
\begin{equation}\label{B.23}
x_n = \Sigma'(t_1) = \int_{t_0}^{t_1}\!\ud\Sigma'(t) > 0
\end{equation}
where we use \eqref{B.12a}.
From \eqref{B.21} we can represent $\nvec{x}$ as
\begin{equation}\label{B.24}
\nvec{x} = x_n\, \nvec{x}'
\end{equation}
where
\begin{gather}
\nvec{x}' = \int_{t_0}^{t_1}  \begin{pmatrix}
    y_1(t)/y_n(t)\\ \vdots \\ y_{n-1}(t)/y_n(t) \\ 1
  \end{pmatrix} \ud\Sigma''(t)\,,
\notag\\ \label{B.25}
\ud\Sigma''(t) = \frac{1}{x_n}\ud\Sigma'(t)\,,\qquad
\int_{t_0}^{t_1}\!\ud\Sigma''(t) = 1\,.
\end{gather}
Clearly $\nvec x'$ is in the intersection of $\coconecl(L')$ with the
hyperplane $x'_n=1$.
Since $\coconecl(L')$ is the minimal closed convex cone containing $L'$ this
intersection is the minimal closed convex set containing $L'$, that is, the closed
convex hull $\cohullcl(L')$:
\begin{equation}\label{B.26}
\nvec{x}' \in \cohullcl(L')\,.
\end{equation}
Thus we have shown that every non-zero vector $\nvec x\in\coconecl(L)$,
that is of the form \eqref{B.19}, can be represented as $x_n\,\nvec x'$ where
$\nvec x' \in \cohullcl(L')$.

Now we come to the application of the above mathematical theorems to our
problems.
Consider three structure functions as in \eqref{f2threeQ2vec} but
-- for simplicity -- only for fixed massless flavor $q$
\begin{equation}\label{B.27}
\begin{pmatrix} F_2^{(q)}(W,Q_1^2)\\ F_2^{(q)}(W,Q_2^2)\\ F_2^{(q)}(W,Q_3^2)
\end{pmatrix} =
\int_0^\infty\! \ud r\, r \frac{\sigmadip^{(q)}(r,W)}{2\pi\alpha_{\text{em}}}
\begin{pmatrix} f^{(q)}(r,Q_1^2)\\ f^{(q)}(r,Q_2^2)\\ f^{(q)}(r,Q_3^2)
\end{pmatrix}.
\end{equation}
To bring this into the form of the problem \eqref{B.5}, \eqref{B.6},
we change variables and set
\begin{equation}\label{B.28}
r = r_0 \frac{t}{1-t}\,,\qquad r_0 = 1~\text{fm}\,,\qquad 0 < t < 1\,.
\end{equation}
Furthermore, we shall split off from $f^{(q)}(r,Q^2)$ the asymptotic terms
for $r\to\infty$ and $r\to 0$.
In the present case of a massless flavor $q$,
$f^{(q)}(r,Q^2)$ decreases for $r\to\infty$ as $1/r^4$, for $r\to 0$ it
behaves as $1/r^2$. Thus, we define a function
\begin{equation}\label{B.29}
g(r) = \frac{r_0^4}{r^2(r_0+r)^2}
\end{equation}
which is independent of $Q^2$. This allows us to define the function
\begin{equation}\label{B.30}
\hat{f}^{(q)}(t,Q^2) =
\begin{cases}
  \lim_{t'\to 0}
  \left.\frac{f^{(q)}(r,Q^2)}{g(r)}\right\vert_{ r = r_0 \frac{t'}{1-t'} }
  & \text{if~} t=0\,,\\
  \phantom{\lim_{t'\to 0}}
   \left.\frac{f^{(q)}(r,Q^2)}{g(r)}\right\vert_{ r = r_0 \frac{t}{1-t} }
  & \text{if~} 0 < t < 1\,,\\
  \lim_{t'\to 1}
  \left.\frac{f^{(q)}(r,Q^2)}{g(r)}\right\vert_{ r = r_0 \frac{t'}{1-t'} }
  & \text{if~} t=1
\end{cases}
\end{equation}
for all $t$ in the closed interval $[0,1]$, since the limites in
\eqref{B.30} exist.
It is easy to show that $\hat{f}^{(q)}(t,Q^2)$ is continuous as function of $t$.
Moreover, we find 
\begin{equation}\label{B.30a}
  0 < c_0(Q^2) \leq \hat{f}^{(q)}(t,Q^2) \leq c_1(Q^2)
\end{equation}
for all $t\in[0,1]$.
Here $c_j(Q^2)$ $(j=0,1)$ are fixed positive constants for fixed $Q^2$.
Next we note that the dipole model makes only sense if the dipole cross
section $\sigmadip^{(q)}(r,W)$ can be integrated with $g(r)$, that is, if
\begin{equation}\label{B.31}
\int_0^\infty\!\ud{r}\,r\,g(r)\,
  \frac{\sigmadip^{(q)}(r,W)}{2\pi\alpha_{\text{em}}}~  < ~\infty\,.
\end{equation}
Further, we assume for $0< r < \infty$:
\begin{equation}\label{B.31a}
\sigmadip^{(q)}(r,W) \geq 0\,.
\end{equation}
This allows us to define a function $\Sigma^{(q)}(t,W)$ which is non-decreasing in $t$ for fixed $W$:
\begin{equation}\label{B.32}
\Sigma^{(q)}(t,W) = \int_0^t\! \ud{t'} \left[
    \frac{\ud{r'}}{\ud{t'}} r'\,g(r')
    \frac{\sigmadip^{(q)}(r',W)}{2\pi\alpha_{\text{em}}}
  \right]_{ r' = r_0 \frac{t'}{1-t'} }
\end{equation}
for $0\leq t\leq 1$.
Conversely, every non-decreasing function $\Sigma^{(q)}(t,W)$ gives, via
\eqref{B.32}, an acceptable dipole cross section $\hat\sigma^{(q)}(r,W)$.
Furthermore, we define the curves $L$ and $L'$ as follows
\begin{alignat}{4}
\label{B.33}
&L:\quad& t &\mapsto &\nvec{y}(t) &:= \begin{pmatrix}
  \hat{f}^{(q)}(t,Q_1^2)\\ \hat{f}^{(q)}(t,Q_2^2)\\ \hat{f}^{(q)}(t,Q_3^2)
  \end{pmatrix}\,,\qquad
  &&0 \leq t \leq 1\,,
\\
\label{B.34}
&L':\quad& t &\mapsto &\nvec{y}'(t) &:= \begin{pmatrix}
  y_1(t)/y_3(t) \\ y_2(t)/y_3(t)\\ 1
  \end{pmatrix}\,,\qquad
  &&0 \leq t \leq 1\,.
\end{alignat}
Our original integrals \eqref{B.27} take now exactly the form of \eqref{B.6}:
\begin{equation}\label{B.35}
\nvec{x} \equiv \begin{pmatrix}
  F_2^{(q)}(W,Q_1^2)\\ F_2^{(q)}(W,Q_2^2)\\ F_2^{(q)}(W,Q_3^2)
  \end{pmatrix} =
\int_0^1\! \nvec{y}(t)\,\ud{\Sigma^{(q)}(t,W)}\,.
\end{equation}
The vector function $\nvec{y}(t)$ is continuous for $t\in[0,1]$ and
\eqref{B.5a} is satisfied with $\nvec{a}^{\mathrm{T}}=(0,0,1)$ due to
\eqref{B.30a} which also guarantees \eqref{B.12a}.
From \eqref{B.8} we conclude that $\nvec x$ must be in the smallest closed
convex cone containing $L$, that is, in $\coconecl(L)$.
It is easy to see that this cone coincides with the cone defined as in
\eqref{f2convexcond} but for fixed flavor $q$.
Indeed, we have from \eqref{B.4f} that $\coconecl(L)$ can be represented as
\begin{equation}\label{B.35a}
\coconecl(L) = \left\{\,\nvec{x}\,\vert~ \nvec{x} = \lambda \nvec{z},~
  \lambda \geq 0,~ \nvec{z}\in\cohullcl(L)\,\right\}.
\end{equation}
With a simple rescaling by $g(r)$ from \eqref{B.29} and using that the
closure takes care of the limiting points $t=0$ and $t=1$ which correspond
to $r\to 0$ and $r\to\infty$ we get for every $\nvec{x}\in\coconecl(L)$ the
representation
\begin{align}\label{B.35b}
&\nvec{x} = \mu\,\nvec{v},\notag\\
&\text{with}\quad
  \mu \geq 0,\quad
  \nvec{v} \in \cohullcl\left\{\left. \begin{pmatrix}
      f^{(q)}(r,Q_1^2)\\ f^{(q)}(r,Q_2^2)\\ f^{(q)}(r,Q_3^2)
    \end{pmatrix}\,\right\vert\,r\in\realnums^+\right\}.
\end{align}
The extension of these arguments to more than one flavor is straightforward.
With this we have given a rigorous proof of \eqref{f2convexcond}.

Consider next the ratio vector
\begin{equation}\label{B.36}
\nvec{x}'= \frac{1}{x_3}\nvec{x} = \begin{pmatrix}
  F_2^{(q)}(W,Q_1^2) / F_2^{(q)}(W,Q_3^2)\\
  F_2^{(q)}(W,Q_2^2) / F_2^{(q)}(W,Q_3^2)\\
  1 \end{pmatrix}.
\end{equation}
From \eqref{B.19}, \eqref{B.24} and \eqref{B.26} we see that
$\nvec x'$ must be in the closed convex hull
$\cohullcl(L')$: 
\begin{equation}\label{B.37}
\nvec{x}' \in \cohullcl(L')\,.
\end{equation}
We have from \eqref{B.34}, \eqref{B.33} and \eqref{B.30}
\begin{align}\label{B.38}
\cohullcl(L')&=\cohullcl\left\{\left. \begin{pmatrix}
  \hat{f}^{(q)}(t,Q_1^2) / \hat{f}^{(q)}(t,Q_3^2)\\
  \hat{f}^{(q)}(t,Q_2^2) / \hat{f}^{(q)}(t,Q_3^2)\\
  1 \end{pmatrix}\,\right\vert\, 0 \leq t \leq 1 \right\}
\notag\\
&=\cohullcl\left\{\left. \begin{pmatrix}
  f^{(q)}(r,Q_1^2) / f^{(q)}(r,Q_3^2)\\
  f^{(q)}(r,Q_2^2) / f^{(q)}(r,Q_3^2)\\
  1 \end{pmatrix}\,\right\vert\, 0 < r < \infty \right\}.
\end{align}
Taking the closure eliminates differences which could otherwise exist
between the two convex hulls of the sets in \eqref{B.38} originating from
the fact that $\hat{f}^{(q)}(t,Q^2)$ is defined on a closed $t$ interval
whereas $f^{(q)}(r,Q^2)$ is defined on an open $r$ interval.
The straightforward extension of \eqref{B.37} and \eqref{B.38} to the case of
several flavors $q$ proves \eqref{f2corbnds}.

With this we have illustrated for one particular case how our bounds are
derived in a mathematically rigorous way.
For all other cases analogous arguments can be applied.

\bibliography{references}

\end{document}